\documentclass[aps,pra,twocolumn,groupedaddress]{revtex4-1}

\usepackage{graphicx,bbm,bm,hyperref}  
\usepackage{amsmath}
\usepackage{amsthm}

\usepackage[pdftex]{color}

\def\ii{{\rm i}}
\def\sx#1{\sigma^{\rm x}_{#1}}
\def\sy#1{\sigma^{\rm y}_{#1}}
\def\sz#1{\sigma^{\rm z}_{#1}}

\def\Z2{\mathbbm{Z}_2}

\def\tr#1{{\rm tr}(#1)}
\def\1{\mathbbm{1}}
\def\ket#1{{| #1 \rangle}}

\def\bracket#1#2#3{{\langle #1|#2 | #3 \rangle}}

\def\tit#1{{\em #1},}
\def\etal#1{#1}
\def\Sup{{Appendix }}

\begin{document}

\title{Weak integrability breaking: chaos with integrability signature in coherent diffusion}

\author{Marko \v Znidari\v c}
\affiliation{Physics Department, Faculty of Mathematics and Physics, University of Ljubljana, 1000 Ljubljana, Slovenia}

\date{\today}

\begin{abstract}
We study how perturbations affect dynamics of integrable many-body quantum systems, causing transition from integrability to chaos. Looking at spin transport in the Heisenberg chain with impurities we find that in the thermodynamic limit transport gets diffusive already at an infinitesimal perturbation. Small extensive perturbations therefore cause an immediate transition from integrability to chaos. Nevertheless, there is a remnant of integrability encoded in the dependence of the diffusion constant on the impurity density, namely, at small densities it is proportional to the square root of the inverse density, instead of to the inverse density as would follow from Matthiessen's rule. We show that Matthiessen's rule has to be modified in non-ballistic systems. Results also highlight a nontrivial role of interacting scattering on a single impurity, and that there is a regime where adding more impurities can actually increase transport.
\end{abstract}





\maketitle

Integrable systems form one of the cornerstones on which our understanding of nature rests. Their solvability leads to an enhanced understanding of that particular system, while on the other hand often enough such simplified models do actually describe realistic systems with a sufficient precision. An example is physics at low energies where description in terms of non- or weakly-interacting quasiparticles often applies, and if on top of that the ``environmental'' effects are small, one has a perfect experimental test bed of integrable physics. The last decade has seen a broad expansion of interest to genuine many-body systems with interactions that are not integrable and to generic high energy states. A pertinent question is, how, if at all, is integrability that is often only weakly broken, reflected in properties of a non-integrable model as probed in an out of equilibrium situation~\cite{exper,noneq}?

We study two questions: (i) breaking of integrability in a many-body system and, in particular, at what perturbation strength does one get a full generic complexity associated with ergodicity, decay of correlations and in our case diffusive transport, and (ii) after integrability is broken and transport goes from non-diffusive (typical of integrable systems~\cite{prelovsek97}) to diffusive, is there some remaining signature of the parent integrability, or it vanishes completely, making integrable systems an utterly singular notion that immediately goes into ``featureless'' diffusion in the thermodynamic limit (TDL)? We find that the critical perturbation strength for the transition from integrability to chaos is zero in the TDL. Nevertheless, the original integrability is still reflected in a modified Matthiessen's rule -- in general the diffusion constant is not simply inversely proportional to the density of impurities.

From a single-particle quantum chaos~\cite{Haake}, or few-degrees-of-freedom classical systems, we know that the transition from integrability to chaos typically happens at a finite perturbation strength (for classical systems the KAM theorem makes that rigorous~\cite{Gutzwiller}). For many-body quantum systems one might expect that the transition strength will instead go to zero in the TDL, results though are not always as clear cut despite a long history, e.g.~\cite{poilblanc93,hsu93,dima97,prosen98}. For instance, while traditional criteria of single-particle quantum chaos like the nearest-neighbor level spacing distribution (LSD) typically do show a transition at zero perturbation strength~\cite{hsu93,dima97,rigol10} in the TDL, looking at the decay of correlation functions there are observations of non-ergodicity at finite perturbations~\cite{prosen98}. An important point to keep in mind is that the LSD probes unobservable exponentially small energy scales and is not always a suitable indicator of complexity (chaos). For instance, a small local perturbation suffices to make a system ``chaotic'' according to the LSD~\cite{Lea,Brenes,Brenes20,Lea20}, despite transport remaining that of an integrable model (ballistic)~\cite{Brenes}. Coexistence of ``chaotic'' LSD and non-ergodic wavefunctions can be observed also in disordered systems~\cite{altland19,anto14}. It is therefore important to better understand integrability to chaos transition in many-body systems in terms of observables in as large systems as possible in order to correctly account for long time- and length-scales emerging at weak perturbations, a problem which can plague exact diagonalization studies of the transition.

\begin{figure}[b!]
\vskip2mm
\centerline{\includegraphics[width=.75\columnwidth]{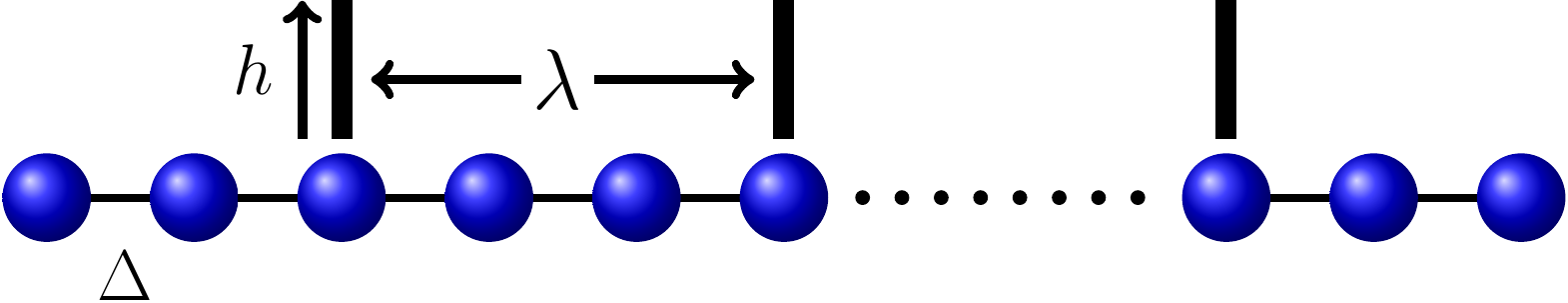}}
\caption{(Color online) XXZ chain (\ref{eq:H}) with magnetic field of amplitude $h$ at sites separated by distance $\lambda$ (shown is $\lambda=3$).}
\label{fig:chain}
\end{figure}
We do that by studying transport in the Heisenberg spin-$1/2$ chain with integrability-breaking impurities (Fig.~\ref{fig:chain}). The model is appealing for a number of reasons. (i) Without impurities it is integrable, with spin transport at high temperature well understood. (ii) The chosen perturbation allows us to study three different kinds of integrability breaking: the interaction $\Delta$, the impurity strength $h$, and the impurity density $1/\lambda$. (iii) The model is experimentally relevant, realized in a number of spin-chain materials like strontium cuprate, where high heat conductivity measured at low-$T$ is attributed to ballistic spin transport along Heisenberg chains~\cite{Hess19}. Because crystals are never perfect~\cite{Hlubek12}, or by deliberately introducing impurities~\cite{Kawamata08,Hlubek11}, one in fact always deals with the Heisenberg model with low density of impurities -- precisely what we study. Transport in the Heisenberg model has also been studied in cold-atoms experiments~\cite{Fukuhara13,Hild14,Ketterle20} and with neutron scattering~\cite{Moore}, promising an even greater controllability in the future. (iv) Importantly, transport at an infinite-$T$ can be studied in large systems, avoiding finite-size effects.

\begin{figure}[t!]
\centerline{\includegraphics[width=.6\columnwidth]{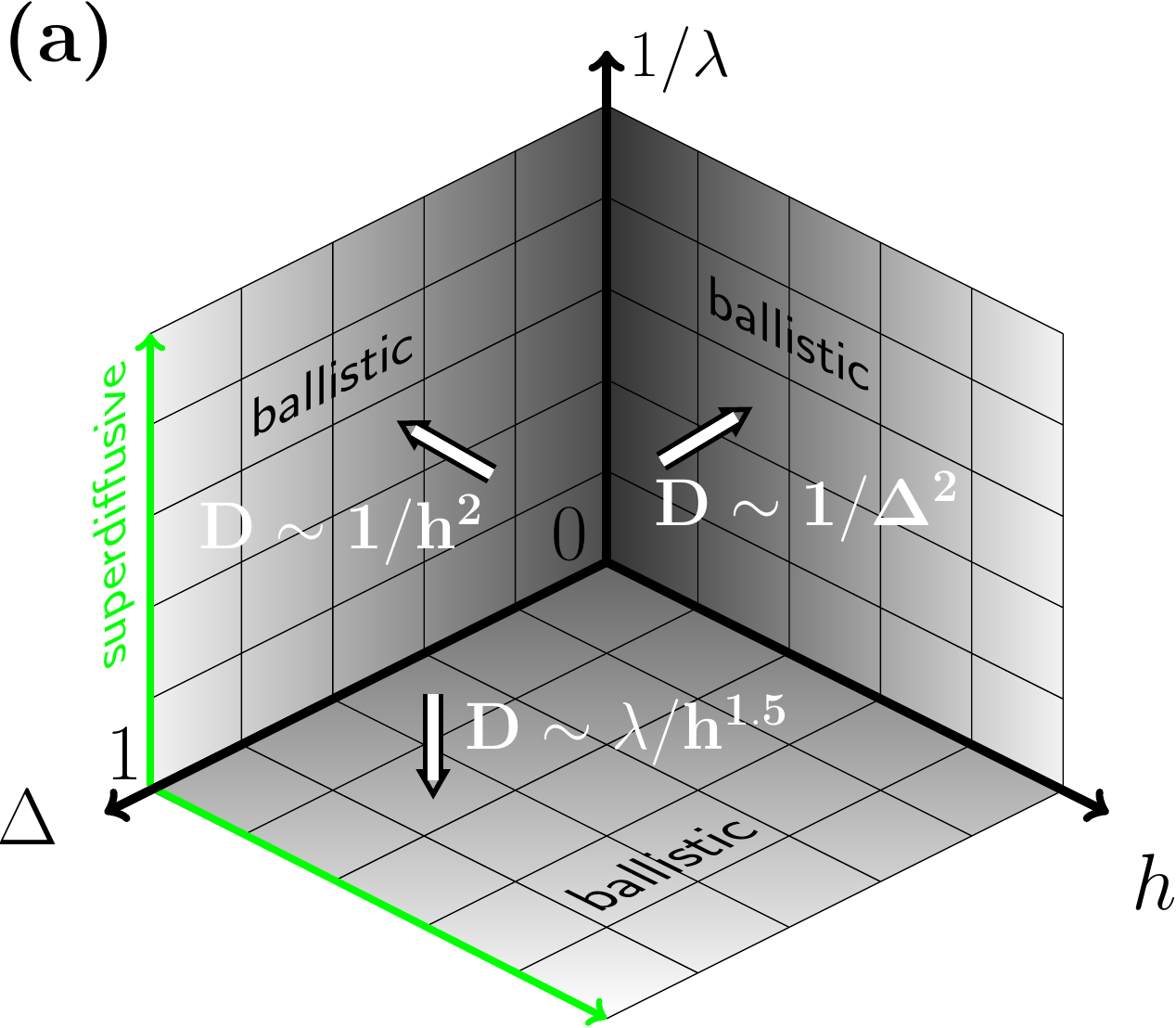}\includegraphics[width=.4\columnwidth]{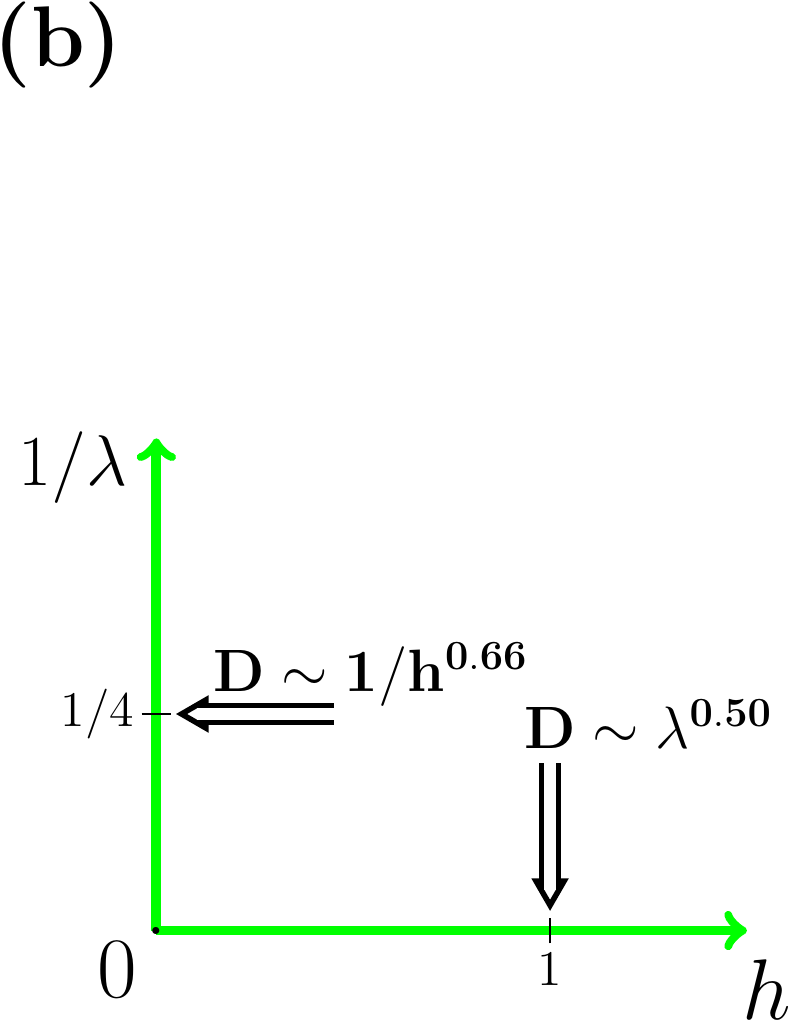}}
\caption{(Color online) Summary: (a) When any small parameter $h, \Delta$ or $1/\lambda$  is $0$ one has known ballistic transport. For nonzero perturbations one gets diffusion with white arrows indicating how diffusion constant $D$ diverges. (b) $\Delta=1$, where one has superdiffusion (green) without perturbation.}
\label{fig:shema}
\end{figure}

What we find is that the faster-than-diffusive spin transport of the integrable model goes upon integrability-breaking immediately to diffusion, with a diverging diffusion constant $D$ at small perturbations (see Fig.~\ref{fig:shema}). For dilute impurities, $\lambda \gg 1$, one would expect $D \propto \lambda$ because the scattering on different impurities is independent, making the rates $1/\tau_i$ additive, leading under a simple kinetic Drude formula $D \sim v^2 \tau$ to $D \propto \lambda$ -- the famous Matthiessen's rule~\cite{Kittel} that is indeed observed in the mentioned Heisenberg spin-chain materials~\cite{Hlubek11} or, e.g., dilute alloys~\cite{Kittel}. What we find, however, is that Matthiessen's rule has to be modified to $D \propto \lambda^{2-z}$, where $z$ is the dynamical transport exponent of the integrable model ($z=\frac{3}{2}$ for the superdiffusive isotropic Heisenberg chain at $T=\infty$). We also find other intriguing features: for $\Delta<1$ and large $\lambda$ the diffusion constant has a nontrivial dependence on $h$ that can be explained by interacting scattering on a single impurity, and there is a regime of high impurity density where spin transport gets faster upon increasing the number of impurities.

Because we focus on transport that is defined in the TDL $\lim_{t \to \infty} \lim_{L \to \infty}$ we do not directly probe finite-time behavior. However, one can note that the way $D$ diverges for small perturbations is indicative of relaxation timescales. We therefore expect that the physics we find in $D$ should be also reflected in finite-time phenomena like prethermalization~\cite{prethermalization}. Another approach dealing with near-integrable systems is using generalized hydrodynamics~\cite{bruno,doyon} and/or conserved quantities to study dynamics upon weak integrabilty breaking~\cite{zala17,Cao18,Caux19,Friedman20,Doyon,Vasseur}.

{\em \bf Results.--}
The anisotropic Heisenberg spin-1/2 chain~\cite{Heisenberg:28} with periodic impurities is
\begin{equation}
H=\sum_{r=0}^{L-1} \sx{r}\sx{r+1}+\sy{r}\sy{r+1}+\Delta \sz{r}\sz{r+1}+h\sum_{k=1}^{M=L/\lambda} \sz{\left \lfloor k\frac{L}{M+1} \right \rfloor},
\label{eq:H}
\end{equation}
where $M=\frac{L}{\lambda}$ is the number of impurities, $\lambda$ the distance between them, and $h$ the size of magnetic field (see Fig.\ref{fig:chain})~\cite{foot1}. We shall focus on spin transport at an infinite temperature and zero magnetization (half-filling). Chaos is a property of generic states and so the ensemble with $T=\infty$ is the most unbiased, and at the same time the easiest to simulate with our numerical method. Without impurities the model is integrable, with spin transport at half-filling and $T=\infty$ being ballistic for $\Delta<1$~\cite{Zotos99,Prosen11,enej17}, and superdiffusive at $\Delta=1$~\cite{PRL11,sarang19,Vir20}. Because we will focus on the breaking of this faster-than-diffusive integrable transport to diffusion we shall not consider $\Delta>1$ where it is diffusive already without impurities~\cite{sarang19}. Any nonzero number of impurities makes the model in general non-integrable~\cite{Lea}. We note that with a single impurity (finite $L$ and $\lambda=\infty$) the spin transport is the same~\cite{Brenes} as for the clean integrable model. Previous studies of transport in the Heisenberg model at high-$T$ under various (weak) perturbations include Refs.~\cite{Zotos96,Alvarez02,Fabian03,vadim06,Rosch07,Moore13,Robin15,Robin16,marko20,michele20}.

\begin{figure}[t!]
\centerline{\includegraphics[width=.75\columnwidth]{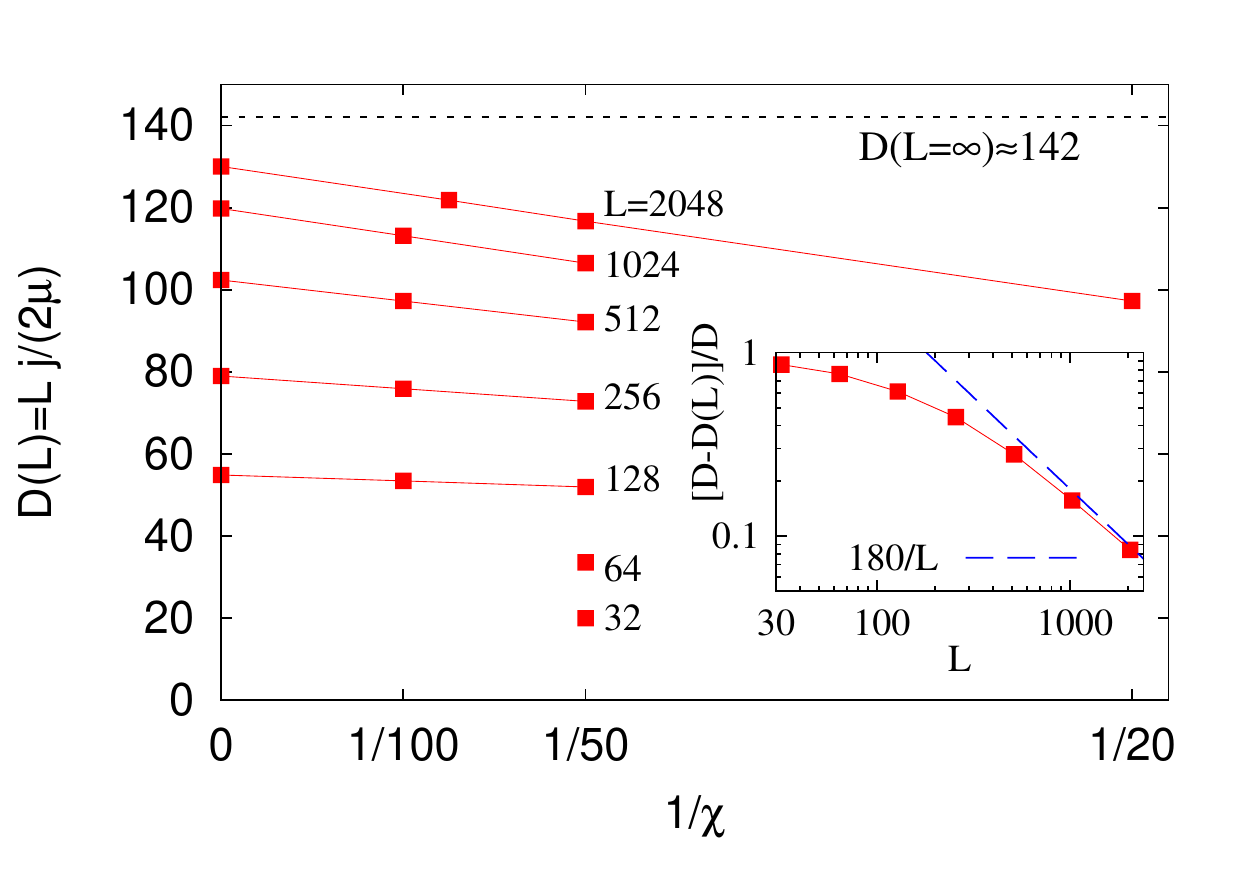}}
\caption{(Color online) Determining diffusion constant ($\lambda=32, \Delta=0.6, h=0.5$). Main plot: convergence with the bond dimension $\chi$ of a finite-size value of $D(L)$, together with the extrapolated values plotted at $1/\chi=0$ (for $L=32,64$ no extrapolation is used). Inset: relative precision of $D(L)$ improves as $\sim 1/L$, as predicted~\cite{nessKubo}, but with a large prefactor $\approx 180$.}
\label{fig:konvD06h05}
\end{figure}
\begin{figure*}[ht!]
\centerline{
\includegraphics[width=.66\columnwidth]{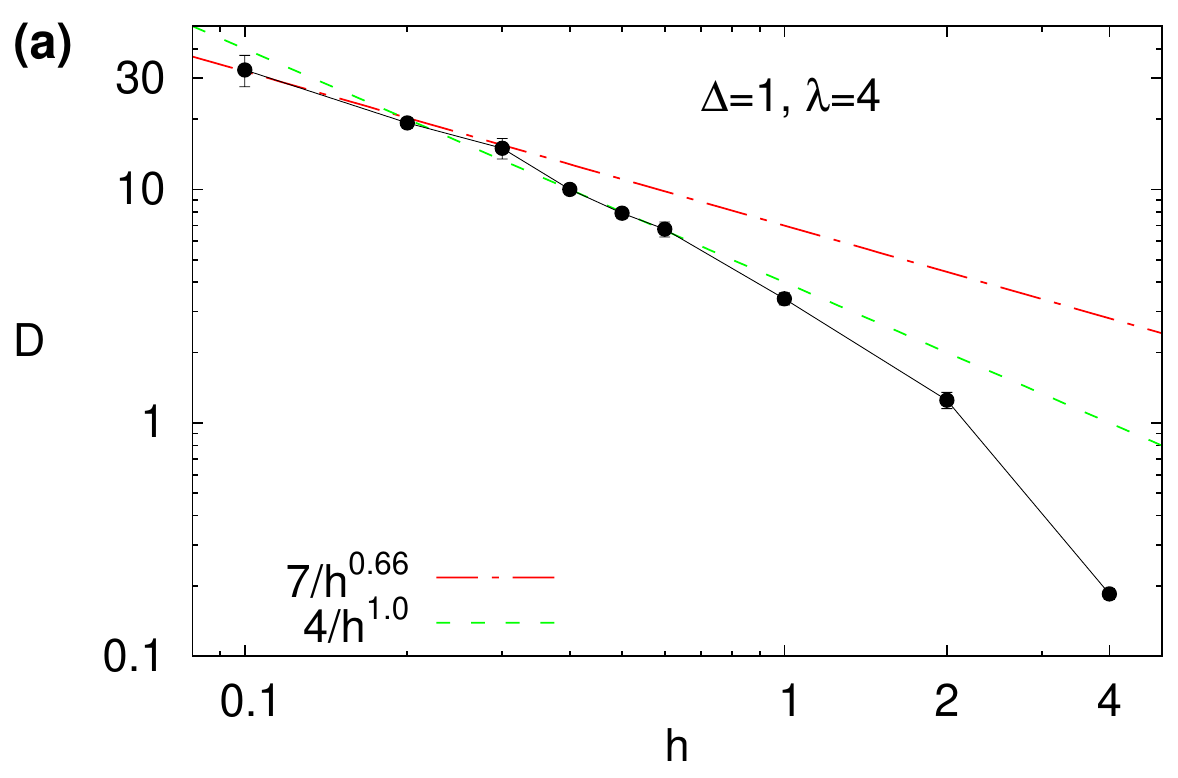}
\includegraphics[width=.63\columnwidth]{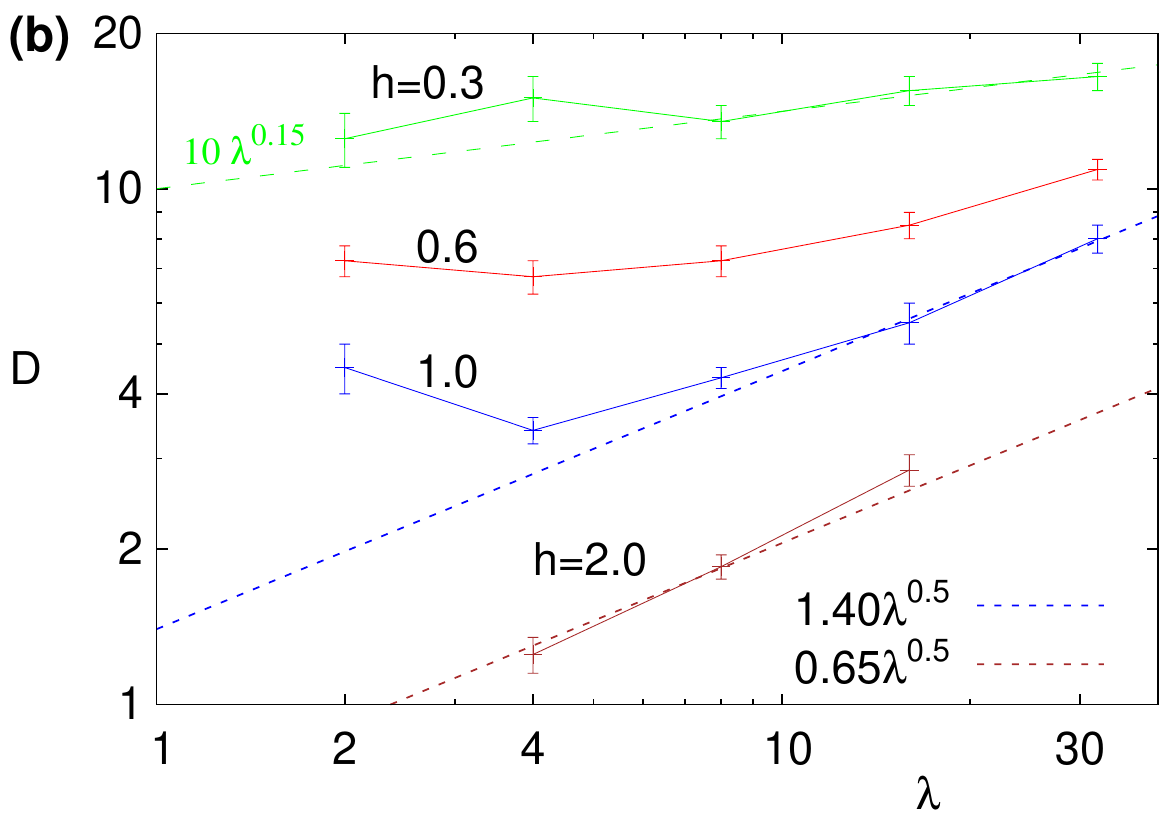}
\includegraphics[width=.63\columnwidth]{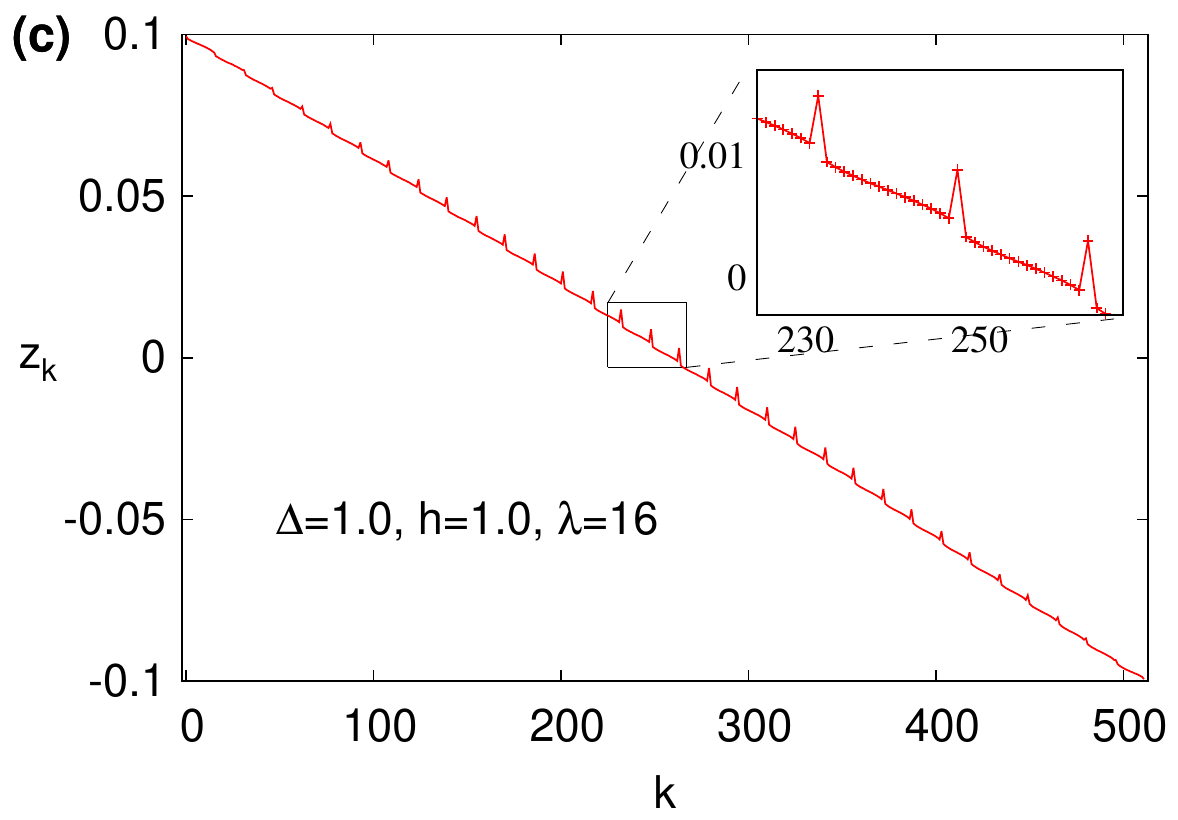}
}
\caption{(Color online) Isotropic chain, $\Delta=1$. (a) Dependence of $D$ on $h$ for $\lambda=4$. (b) Diffusion constant scaling with $\lambda$; for $\lambda \gg 1$ it is not proportional to $\lambda$ (inverse impurity density) but is rather $D \sim \lambda^{0.5}$. (c) Magnetization profile has spikes at impurities ($h=1, \lambda=16$, $L=512$).}
\label{fig:XXX}
\end{figure*}
To numerically assess spin transport we are going to couple the spin chain at first and last sites to magnetization reservoirs described by Lindblad operators $L_1=\sqrt{\Gamma(1+\mu)}\,\sigma^+_0, L_2= \sqrt{\Gamma(1-\mu)}\, \sigma^-_0, L_3 =  \sqrt{\Gamma(1-\mu)}\,\sigma^+_{L-1}$ and $L_4= \sqrt{\Gamma(1+\mu)}\, \sigma^-_{L-1}$, such that the evolution of the density matrix is described by the Lindblad master equation~\cite{Lindblad1,Lindblad2}. Its solution converges at long times to a unique nonequilibrium steady state (NESS) whose properties determine transport, in particular the NESS spin current $j=\tr{\rho (2\sx{k}\sy{k+1}-2\sy{k}\sx{k+1} )}$, and magnetization at site $k$, $z_k=\tr{\rho\sz{k}}$. For zero $H$ the chosen Lindblad operators would induce a steady-state $\rho \sim \1+\mu\sz{0}$ on the 1st site, and $\rho \sim \1-\mu\sz{L-1}$ on the last site (independent of $\Gamma$). They therefore try to induce magnetization $+\mu$ and $-\mu$, respectively, and so $2\mu$ can be thought of as the driving potential difference. Nonzero $H$ makes dynamics and the NESS nontrivial, with the transport type being encoded in the dependence of $j$ on $L$, as well as in the shape of the magnetization profile. For diffusive systems in the linear response regime (small $\mu$; we use $\mu=0.1$) the profile will be on average linear (see Fig.~\ref{fig:XXX}c for an example) while the current will scale as $j \asymp -D \frac{2\mu}{L}$, from which one can extract the diffusion constant $D$. At $\mu=0$ the NESS is a trivial $\sim \mathbbm{1}$ corresponding to an equilibrium $T=\infty$ driving. At $\mu \ll 1$ the NESS is still close to $\1$, energy density is zero, and so the driving probes transport at $T=\infty$ and at zero average magnetization. The coupling strength $\Gamma$, which only influences the boundary resistance, is set to $\Gamma=1$ (see \Sup for more details on $\mu$ and $\Gamma$). Note that the particular choice of driving does not influence the bulk transport properties, specifically, the extracted diffusion constant is the same as the one obtained from the Green-Kubo approach~\cite{nessKubo}.

To represent a solution of the Lindblad equation $\rho(t)$ efficiently we use a matrix product operator ansatz with matrices of size $\chi$ and the tDMRG method~\cite{Schollwock} to evolve $\rho(t)$ in time. The method has proved itself in the past, see, e.g., Ref.~\cite{Znidaric16} and references therein for more details, and allows at ``easy'' parameter values to simulate systems as large as $L \approx 2000$ sites. The crucial parameter that determines its efficiency is $\chi$. The largest $\chi$ we can afford is about $\chi \sim 100$ at $L \sim 1000$ ($\chi \approx 300$ for some smaller $L$). For parameters where truncation errors are larger we run simulations at different $\chi$ and use extrapolation to gain in accuracy (Fig.~\ref{fig:konvD06h05}).
 
We first check the isotropic chain, $\Delta=1$. Fixing $\lambda=4$ we calculate the NESS for increasingly smaller values of magnetic field $h$, each time studying the scaling of $j$ with $L$. In all cases we find diffusive $j \sim 1/L$, see \Sup for data. In Fig.~\ref{fig:XXX}a we plot the obtained $D(h)$. According to Fermi's golden rule, the scattering rate should scale as $1/\tau \sim h^2$. In a system with dynamical exponent $z$, defined by the scaling of distance with time as $x^z \sim t$ (and the NESS current as $j \sim 1/L^{z-1}$), e.g., $z=1$ for ballistic, $z=2$ for diffusion, the scattering length should go as $l \sim  1/h^{2/z}$. For the isotropic model at an infinite temperature $z=\frac{3}{2}$~\cite{PRL11}, predicting divergence $D \sim 1/h^{2/3}$, similarly as for a disordered potential~\cite{Znidaric16}. Numerical results in Fig.~\ref{fig:XXX}a agree with that scaling (the agreement is achieved only at very small $h\lesssim 0.3$; at larger $h$ the scaling power is larger). From an experimental point of view we would in particular like to understand the case of dilute impurities, $\lambda \gg 1$. To that end we plot in Fig.~\ref{fig:XXX}b the scaling of $D$ with $\lambda$ for several values of $h$. We see that $D$ is not proportional to $\lambda$. This is due to non-ballistic transport between impurities and can be explained as follows. Focusing on a segment of length $\lambda$ between two impurities, the magnetization difference across the segment is $\delta z \approx 2\mu/M=2\mu\lambda/L$ and will drive the current of size $j \sim \delta z/\lambda^{z-1}$ through the segment. The last relation comes because an excitation needs time $\sim \lambda^z$ to travel across the length $\lambda$, resulting in a current $\sim \lambda/\lambda^z$ (at fixed excitation density there are $\sim \lambda$ excitations in a segment of length $\lambda$). The NESS current therefore scales as $j \sim \frac{2\mu}{L} \frac{\lambda}{\lambda^{z-1}}$, giving
\begin{equation}
  D \sim \lambda^{2-z}.
  \label{eq:Dz}
\end{equation}
Using $z=\frac{3}{2}$ of the isotropic model we see that the resulting $D \sim \sqrt{\lambda}$ agrees within numerical errors with data in Fig.~\ref{fig:XXX}b. Deviations seen for smaller $h=0.6,0.3$ are presumably due to the scattering length being larger than $\lambda=32$, which is the largest $\lambda$ we can reliably simulate. In Fig.~\ref{fig:XXX}c we plot the magnetization profile across a chain, showing nonequilibrium spikes at locations of impurities (spikes are not visible for all parameters, and are typically stronger at smaller $D$).

\begin{figure}[t!]
\centerline{\includegraphics[width=.75\columnwidth]{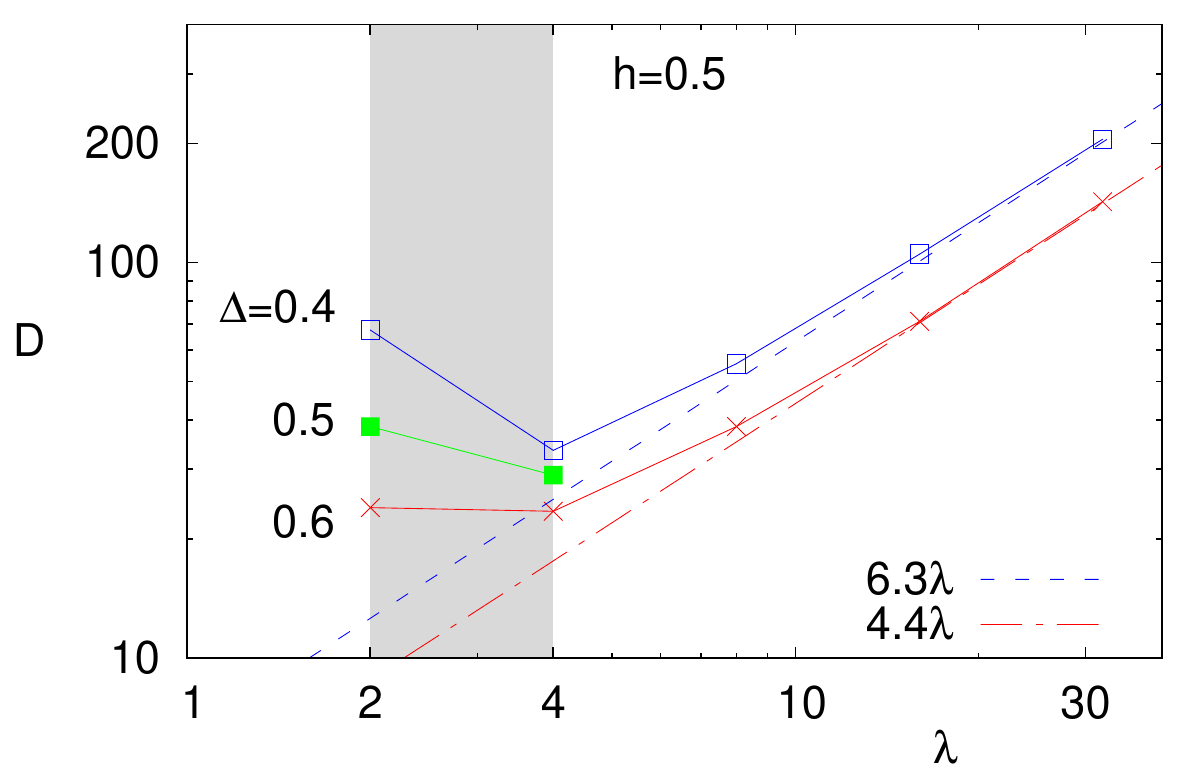}}
\caption{(Color online) Diffusion scaling for $\Delta<1$ ($h=0.5$). For $\lambda \gg 1$ it is linear, with the prefactor given by a single-impurity physics (Fig.~\ref{fig:XXZ}), while in the shaded strip $D$ counterintuitively increases by increasing the number of impurities.}
\label{fig:Dodlam}
\end{figure}
\begin{figure*}[t!]
\centerline{
\includegraphics[width=.65\columnwidth]{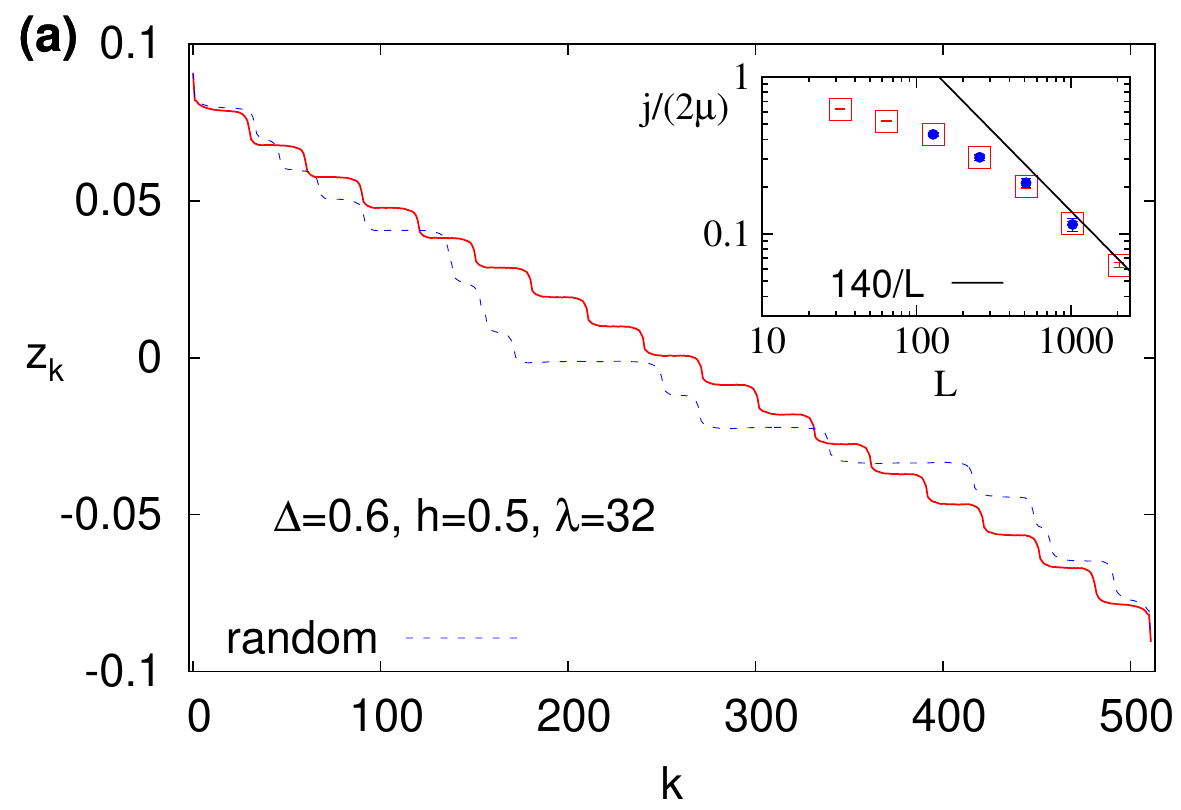}
\includegraphics[width=.65\columnwidth]{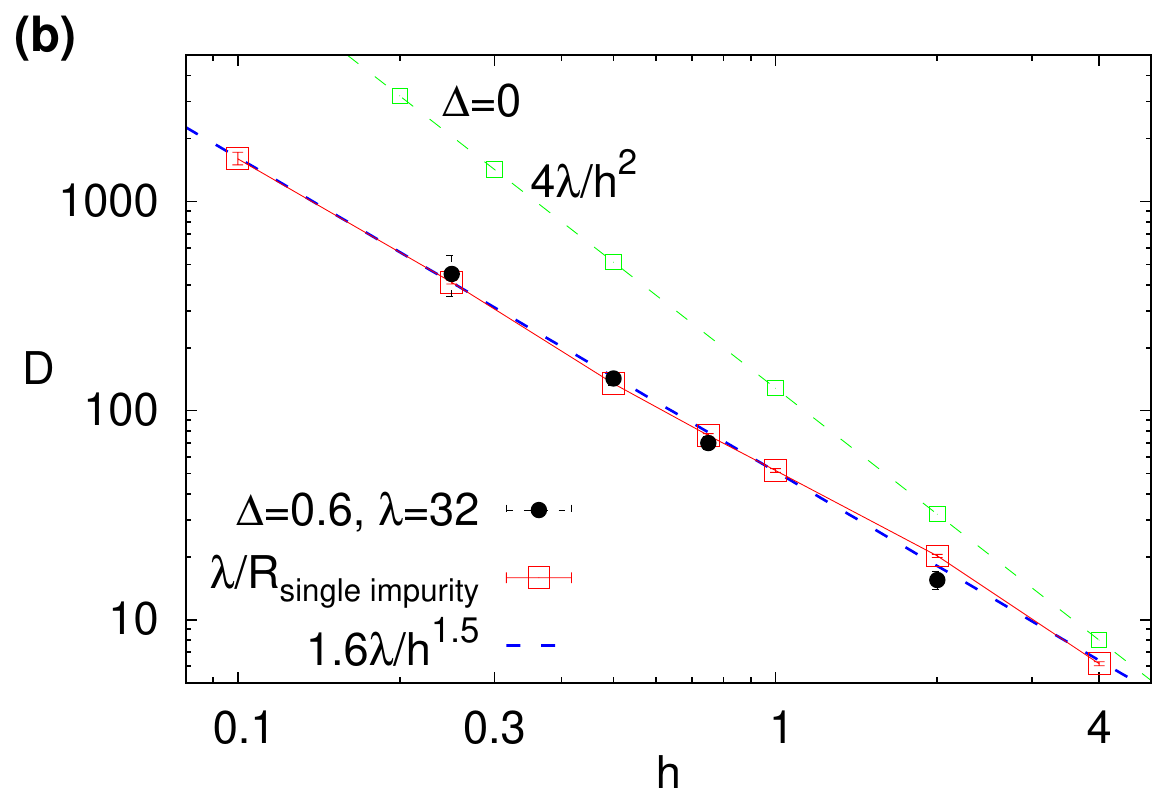}
\includegraphics[width=.63\columnwidth]{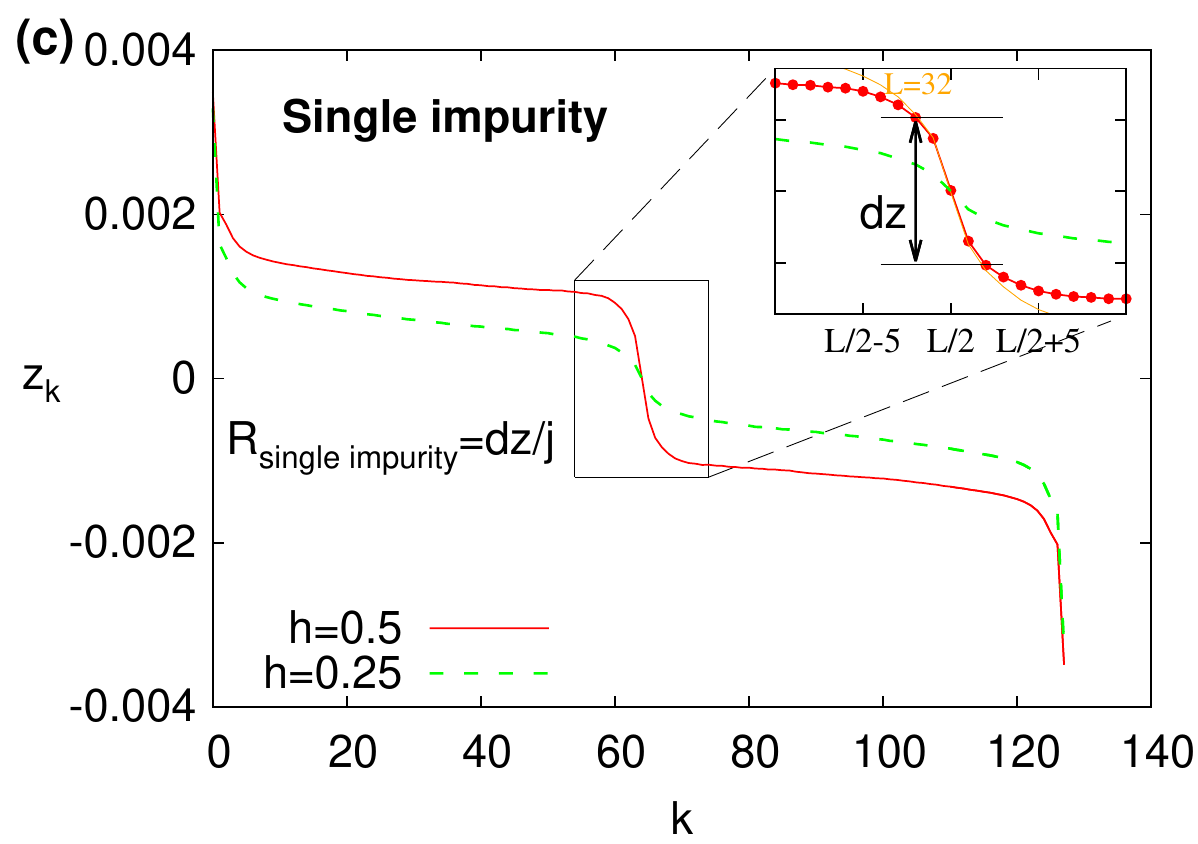}
}
\caption{(Color online) Anisotropic XXZ with $\Delta=0.6$. (a) Magnetization profile for $h=0.5$ and $\lambda=32$ in a chain with $L=512$ (red), and for randomly placed $L/\lambda=16$ impurities (dashed blue). Inset: scaling of the NESS current with $L$ giving $D\approx 140$. Blue points (overlapping with red squares for $\lambda=32$) show the current for the random case. (b) Scaling of $D(h)$ for $\lambda=32$ (4 black circles). Red squares are obtained (no fitting parameters) from the single-impurity scattering in frame (c). Green squares is the exact noninteracting result for $R_{\rm single}$. (c) Magnetization profile for a single impurity at the middle site ($\mu=0.005$). The main plot shows results for $L=128$ (zoom-in also for $L=32$ and $h=0.5$). Magnetization jump at the impurity, $dz:=z_{L/2-2}-z_{L/2+2}$, and the NESS $j$ is used to plot red squares in frame (b).}
\label{fig:XXZ}
\end{figure*}
Next, we focus on $\Delta<1$ where the integrable model is ballistic. Fixing $\lambda=4$, we have two possible small perturbations, either taking small $\Delta$, or small $h$. We again find that small integrability breaking immediately leads to diffusion. For the two perturbation types Fermi's golden rule gives the scattering length $l \sim 1/\Delta^2$, or $l \sim 1/h^2$, leading to diffusion constant divergence $D \sim 1/\Delta^2$, or $D \sim 1/h^2$, respectively. This is inline with numerical data, see \Sup for data. Increasing $\lambda$ at fixed $h$, and using ballistic $z=1$, Eq.~(\ref{eq:Dz}) predicts $D \propto \lambda$ at $\lambda \gg 1$, which agrees with numerics (Fig.~\ref{fig:Dodlam}). What is interesting is the behavior at small $\lambda$. Between $\lambda=2$ and $4$ the diffusion constant increases by decreasing $\lambda$, meaning that the transport gets faster when we add more impurities. The effect is more prominent at small $\Delta$, and was also visible at $h=1$ in the isotropic case (Fig.~\ref{fig:XXX}b). Let us now focus on $\lambda \gg 1$ and in particular on how $D$ depends on parameters. Using the same argument as in deriving Eq.~(\ref{eq:Dz}) we can see that between rare impurities the magnetization profile will be flat, with a jump happening only at impurities (Fig.~\ref{fig:XXZ}a). We also observe that at $\lambda \gg 1$ it does not matter whether impurities are equidistant, like in our simulations, or at random positions -- $D$ is the same in both cases (the same holds at $\Delta=1$). Therefore one should be able to get $D$ just from studying the size of the jump at a single impurity. This is what we do in Fig.~\ref{fig:XXZ}c. Placing the single impurity at the middle of the chain, we study how the jump size $dz$ scales with $h$, and, in particular, how a single-impurity resistance $R_{\rm single}=dz/j$ scales. We determine $dz$ from the 5 central sites around the impurity (for those the profile is independent of $L$ in the TDL). Numerics indicates that $R_{\rm single} \sim h^{1.5}$ at small $h$ (Fig.~\ref{fig:XXZ}b). In the non-interacting case $\Delta=0$ one can solve the corresponding Lindblad equation exactly (following, e.g., Ref.~\cite{EPJB}), obtaining the odd-$L$ NESS values $j=4\mu \frac{\Gamma+1/\Gamma}{(\Gamma+1/\Gamma)^2+h^2}$, $z_{1,\ldots,(L-1)/2-1}=-z_{(L-1)/2+1,\ldots,L-2}=\mu \frac{h^2}{(\Gamma+1/\Gamma)^2+h^2}, z_0=-z_{L-1}=\mu\frac{1+\Gamma^2+h^2}{(\Gamma+1/\Gamma)^2+h^2}, z_{(L-1)/2}=0$, giving $R_{\rm single}(\Delta=0)=\frac{h^2}{2(\Gamma+1/\Gamma)}$ (the scaling of current with $h$ in the single-impurity situation, including at $\Delta=0$, was numerically studied in Ref.~\cite{Brenes}). We see that the power $\approx 1.5$ in $R_{\rm single}$ at $\Delta=0.6$ is different than $2$ obtained at $\Delta=0$. It is also different than the scaling power $D \sim 1/h^{0.66}$ at $\Delta=1$ (see \Sup for data). It very weakly, if at all, depends on $\Delta$ and could therefore be discontinuous at $\Delta=0$ and $\Delta=1$ (see \Sup). Scattering on a single impurity in an interacting wire therefore seems to be qualitatively different than in a non-interacting one; we were not able to obtain the power $\approx 1.5$ using perturbation theory, leaving this as an interesting problem. $R_{\rm single}$ can now be used to calculate the diffusion constant for $\lambda \gg 1$ in a system that is ballistic without impurities (e.g., $\Delta<1$), obtaining
\begin{equation}
  D = \lambda/R_{\rm single}.
  \label{eq:DR}
  \end{equation}
Data in Fig.~\ref{fig:XXZ}b for full many-impurity numerics agree with that well (due to numerical errors the accuracy of the fitted power $1.5$ is about $10\%$). We have an interesting situation where $D$ is very sensitive to having either $\Delta=0$, or $\Delta=1$. Changing the interaction $\Delta$ just a little away from either of the two points changes $D$ drastically. In fact, in the TDL and $\lambda \to \infty$, or $h \to 0$, the relative change is infinite, coming from different scaling of $D$ with $\lambda$ (\ref{eq:Dz}) as well as different scaling of $R_{\rm single}$ with $h$. As an example, taking chaotic model with $\lambda=32$ and $h=0.5$ we can predict that $D$ increases by about tenfold as one changes the interaction from $\Delta=1$ to $\Delta=0.8$.

{\em \bf Conclusion.--} Using transport at an infinite temperature as an indicator we studied the transition from integrability to chaos in the Heisenberg spin chain with impurities. By large-scale numerical simulations of systems with upto $2000$ spins we find that one gets diffusion already for an infinitesimal perturbation strength, in line with a simple Fermi's golden rule. For the important case of dilute impurities we find that the diffusion constant scales as $D \sim \lambda^{2-z}$, where $z$ is the dynamical exponent of the clean integrable model and $\lambda$ the distance between impurities. In particular, for the isotropic Heisenberg model Matthiessen's rule has to be changed to $D \propto \sqrt{\lambda}$, instead of the usual textbook $D \propto \lambda$. Such scaling arises due to a combination of an anomalous coherent propagation between impurities interspersed by scattering events on impurities. One can obtain $D$ by analyzing scattering on a single impurity in an interacting model. Also interesting is that increasing the impurity density from $\frac{1}{\lambda}=\frac{1}{4}$ to $\frac{1}{2}$ can cause diffusion to become faster. $D$ is for $\lambda \gg 1$ very sensitive to being at the isotropic point. We expect our results to hold also at finite (high) temperatures. 

Traditionally, the quantum chaos community has focused on looking for signatures of chaos (generic behavior) -- here we instead find signatures of integrability (rare behavior) in the form of a modified Matthiessen's rule in an otherwise chaotic model. While we studied a particular model and type of impurities, arguments are general and should hold for other dilute perturbations, e.g. bond disorder~\cite{peter10}, and different interacting models with anomalous transport~\cite{marko18,enej20,ziga20}, perhaps even for the Fibonacci model~\cite{fibo19}. Checking the relation (\ref{eq:Dz}) for other conserved quantities, like energy, is also an interesting problem.

I would like to acknowledge support by Grants No.~J1-1698 and No.~P1-0402 from the Slovenian Research Agency, and ERC OMNES (T.~Prosen) for computational resources.

\newpage
\clearpage

\setcounter{equation}{0}
\setcounter{figure}{0}
\setcounter{table}{0}
\makeatletter
\renewcommand{\theequation}{S\arabic{equation}}
\renewcommand{\thefigure}{S\arabic{figure}}

\appendix

\section{Additional data }
\label{app:A}
\begin{figure}[b!]
\centerline{\includegraphics[width=.75\columnwidth]{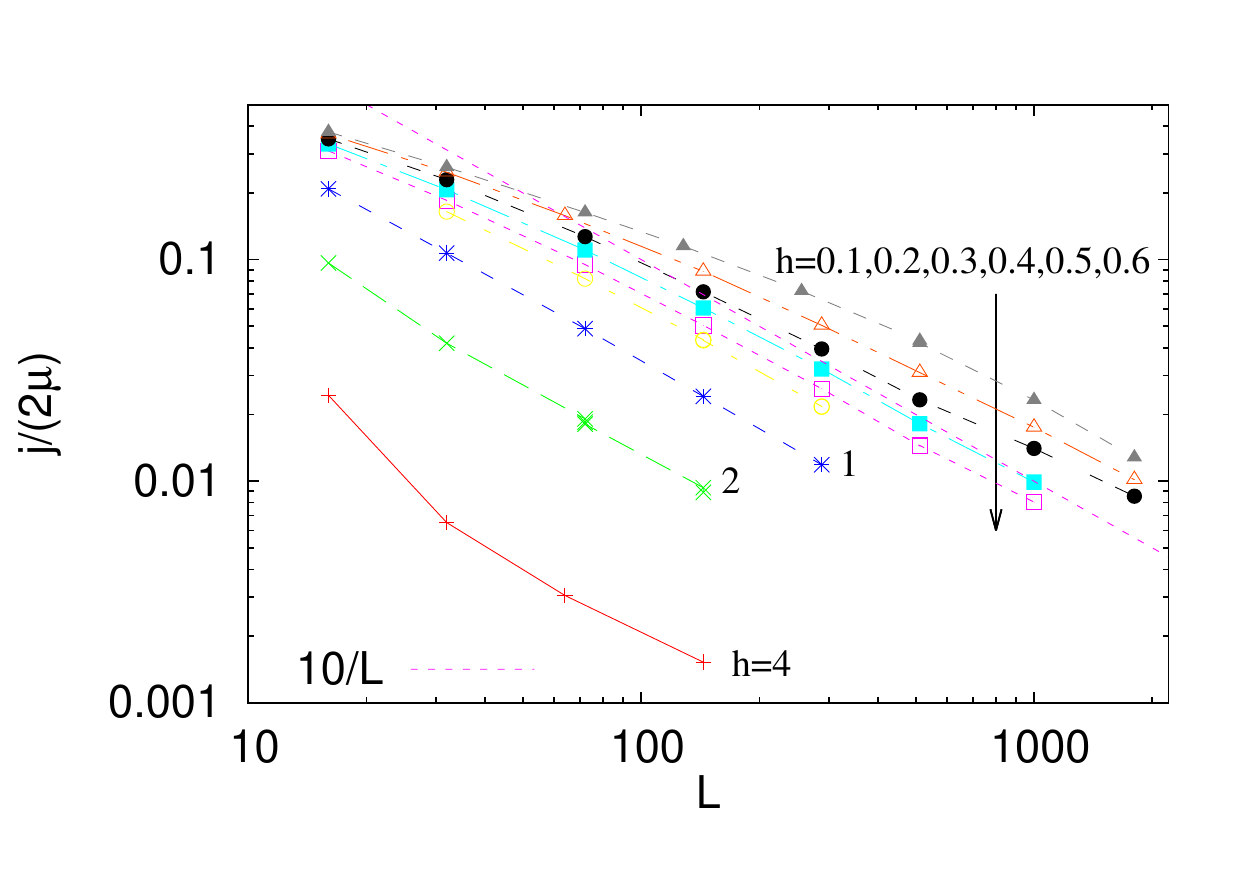}}
\centerline{\includegraphics[width=.75\columnwidth]{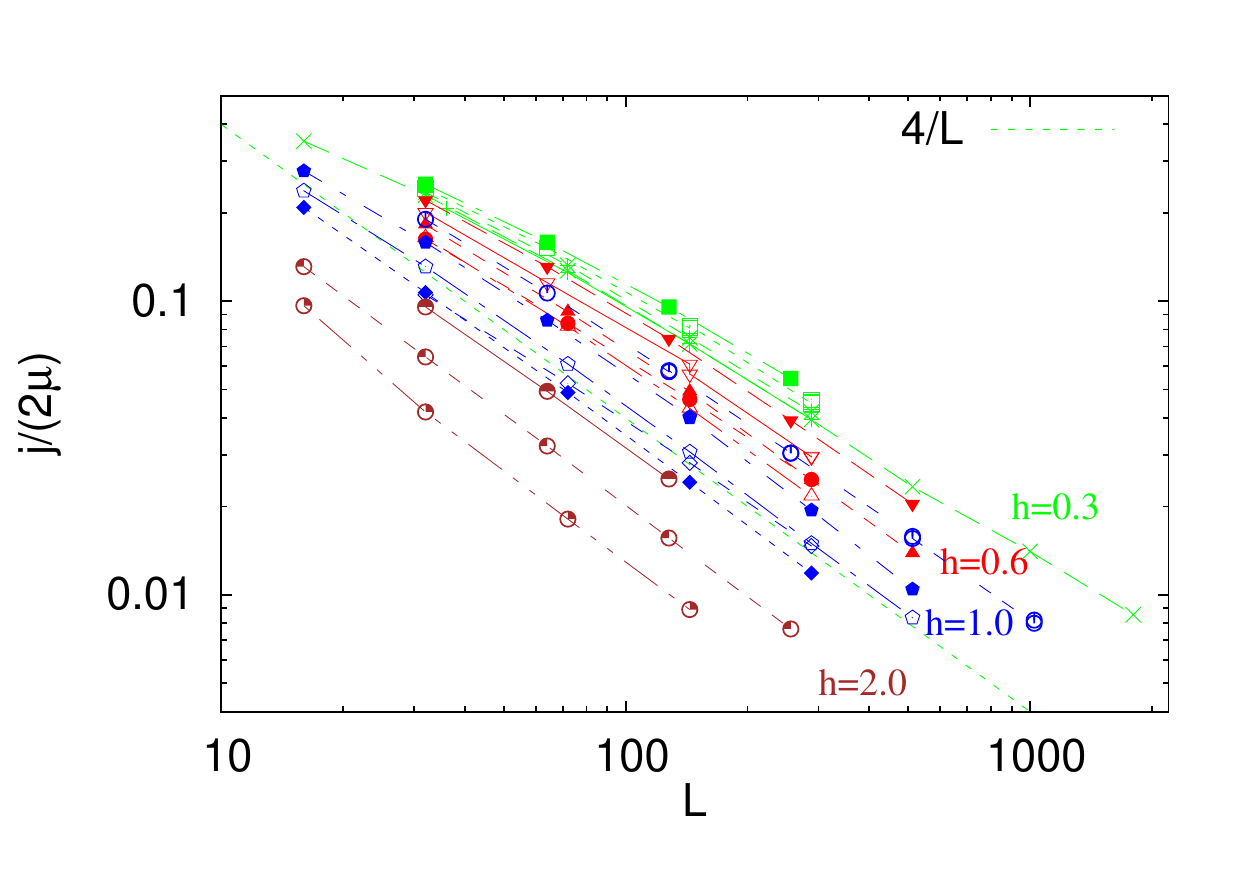}}
\caption{(Color online) Top: Raw data for the isotropic Heisenberg chain and $\lambda=4$ used in Fig.~\ref{fig:XXX}a. Bottom: Similar for different $\lambda$ shown in Fig.~\ref{fig:XXX}b.}
\label{fig:Dall}
\end{figure}

The main object we study is the NESS $\rho$ which is the solution of the stationary Lindblad equation ${\cal L}\rho=0$, with ${\cal L}$ being the Liouvillian generator, i.e., the rhs of the Lindblad equation,
\begin{equation}
\frac{{\rm d}\rho}{{\rm d}t}={\cal L}\rho=\ii[\rho,H]+\sum_k 2L_k \rho L_k^\dagger-\rho L_k^\dagger L_k-L_k^\dagger L_k \rho.
\label{eq:Lin}
\end{equation}
Here we show raw data used to determine diffusion constant $D$. In Fig.~\ref{fig:Dall} we plot the NESS current expectation value $j$ for different $L$ and $h$, all for $\lambda=4$ and $\Delta=1$. In general larger $h$ require larger bond sizes $\chi$, making the method better suited for small perturbations $h$. However, at small $h$ the scattering length is larger and therefore one needs larger systems sizes to get into the asymptotic diffusive regime. To give a rough idea, at $h=4$ we had to use $\chi=200$ for $L=144$ in order to get a bit less than $10\%$ error in the NESS current. On the other hand for $h=0.1$ the bond size $\chi=50$ suffices at $L=100$ to get better than $1\%$ precision, however large sizes are required, and at $L=1800$ we could afford only $\chi=50-80$ at which we estimate the error to be around $10\%$ (a single similar data-point requires about a week or even more of CPU time on $\approx 30$ Xeon cores).

\begin{figure}
\centerline{\includegraphics[width=.75\columnwidth]{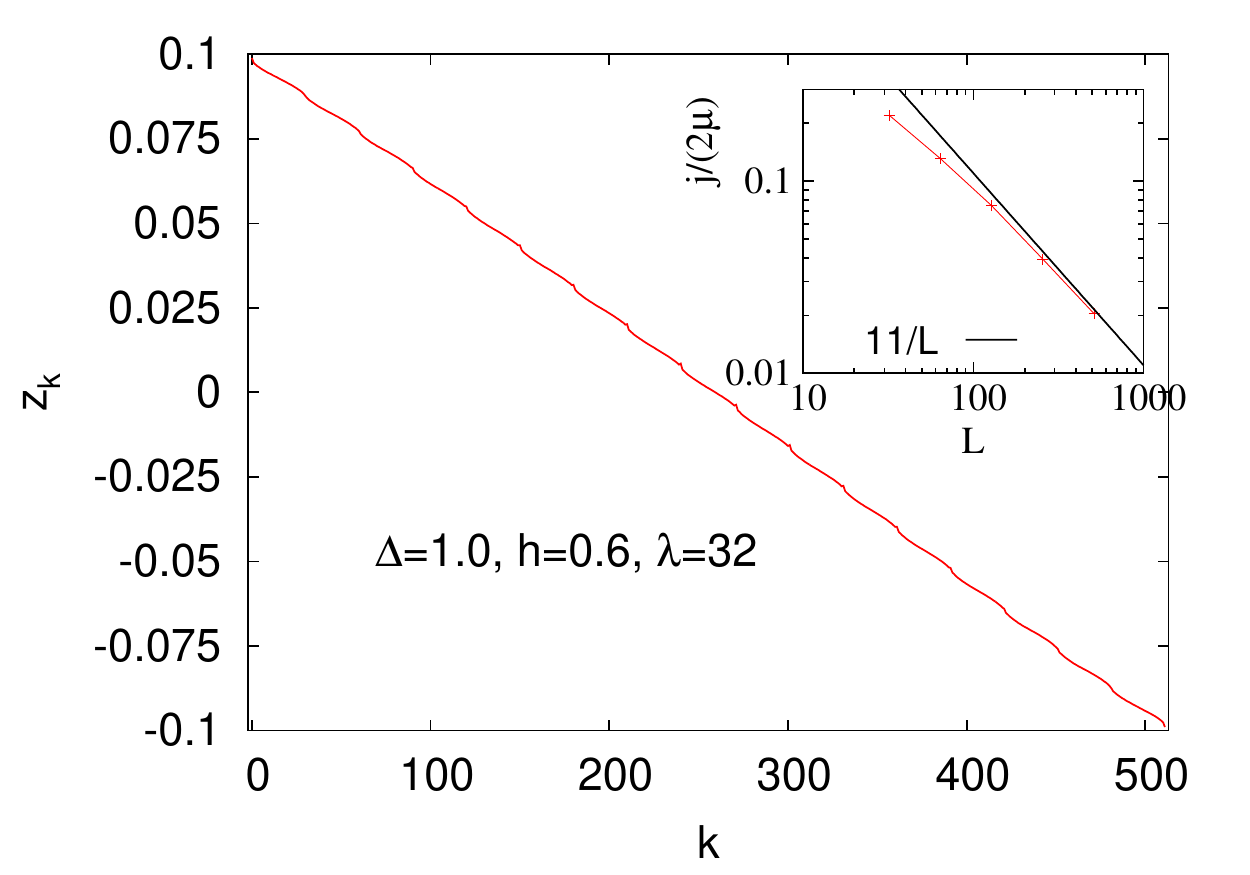}}
\caption{(Color online) Magnetization profile for the isotropic case, $\Delta=1$, $h=0.6, \lambda=32$, $L=512$.}
\label{fig:profilzh06}
\end{figure}

In Fig.~\ref{fig:profilzh06} we show a profile at different parameters than in the main text, where spikes are not as pronounced.

\begin{figure}
\centerline{\includegraphics[width=.75\columnwidth]{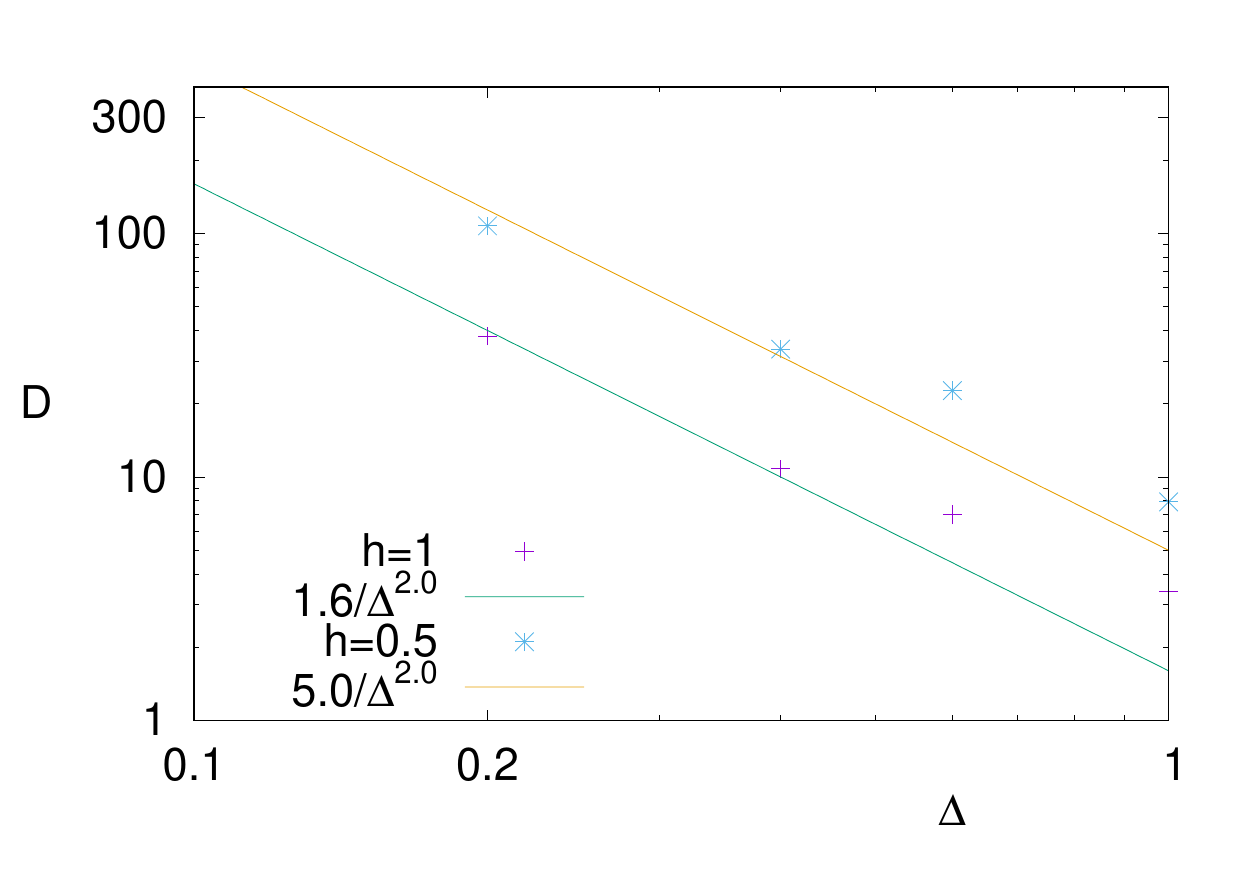}}
\caption{(Color online) For small $\Delta$ one has the expected $D \sim 1/\Delta^2$ ($\lambda=4$).}
\label{fig:DodD}
\end{figure}

Going to the ballistic regime of $\Delta<1$ we checked that the diffusion constant diverges as $\Delta \to 0$, shown in Fig.~\ref{fig:DodD}. We also see that the prefactor $a$ in $D \approx a/\Delta^2$ rapidly increases as $h$ gets smaller.

\begin{figure}
\centerline{\includegraphics[width=.75\columnwidth]{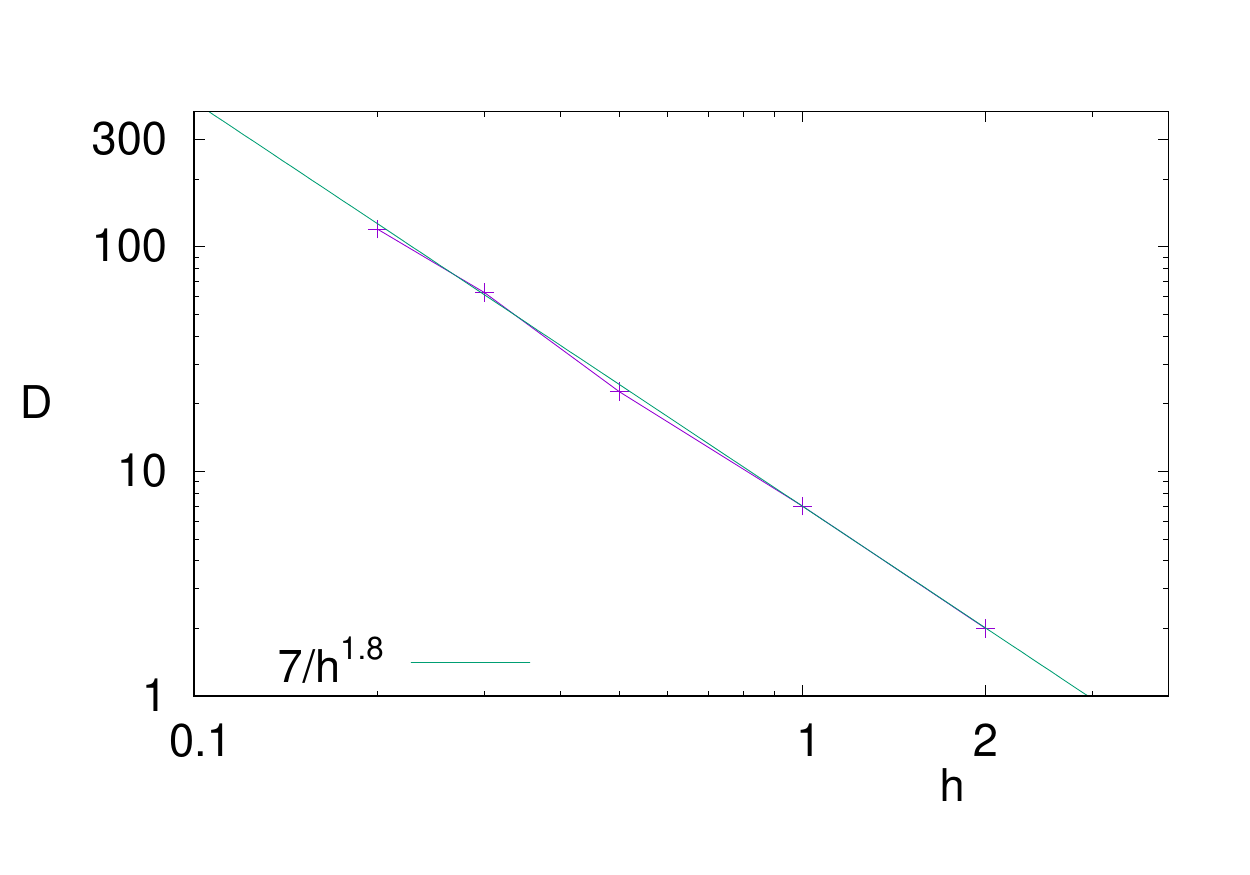}}
\caption{(Color online) Diffusion constant divergence for $\Delta=0.6$ and $\lambda=4$.}
\label{fig:Dodh}
\end{figure}

\begin{figure}
\centerline{\includegraphics[width=.75\columnwidth]{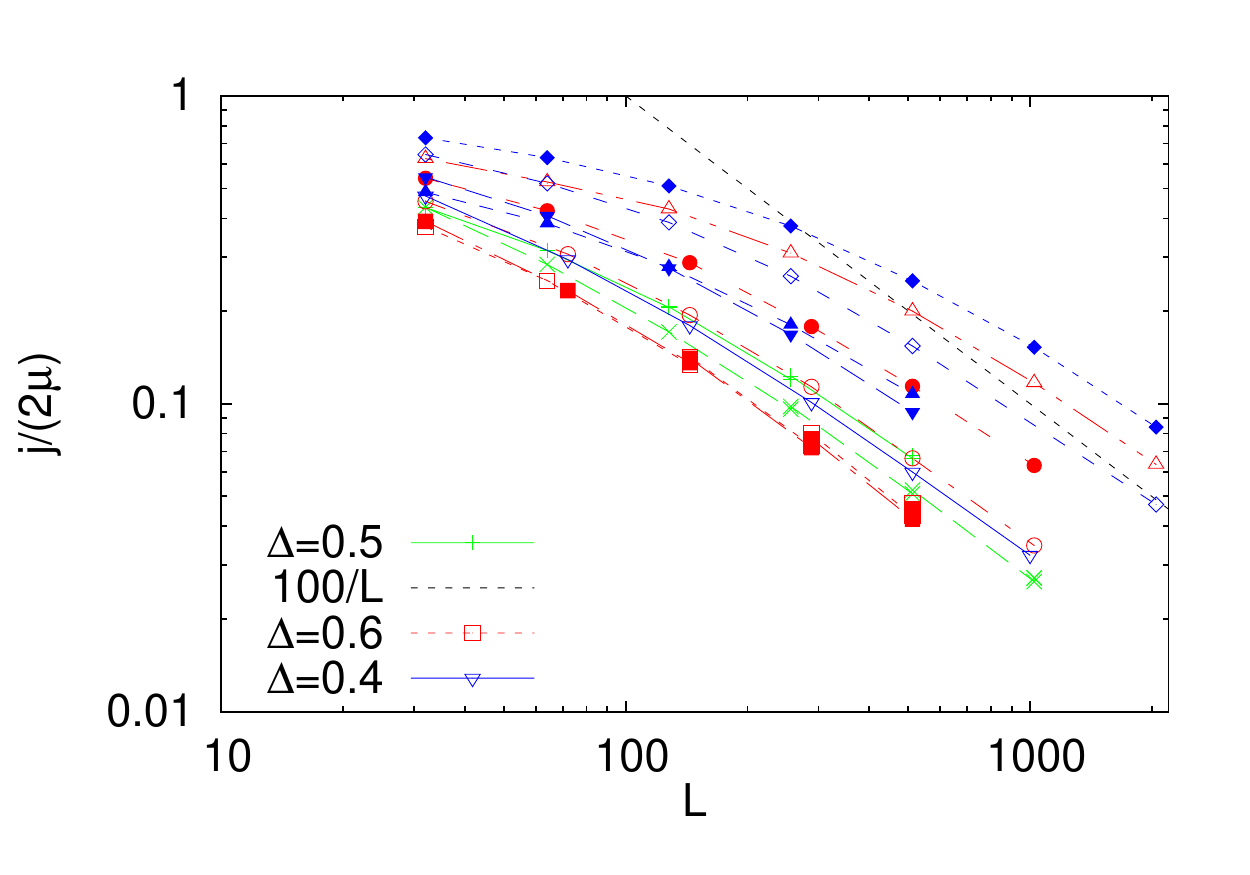}}
\centerline{\includegraphics[width=.75\columnwidth]{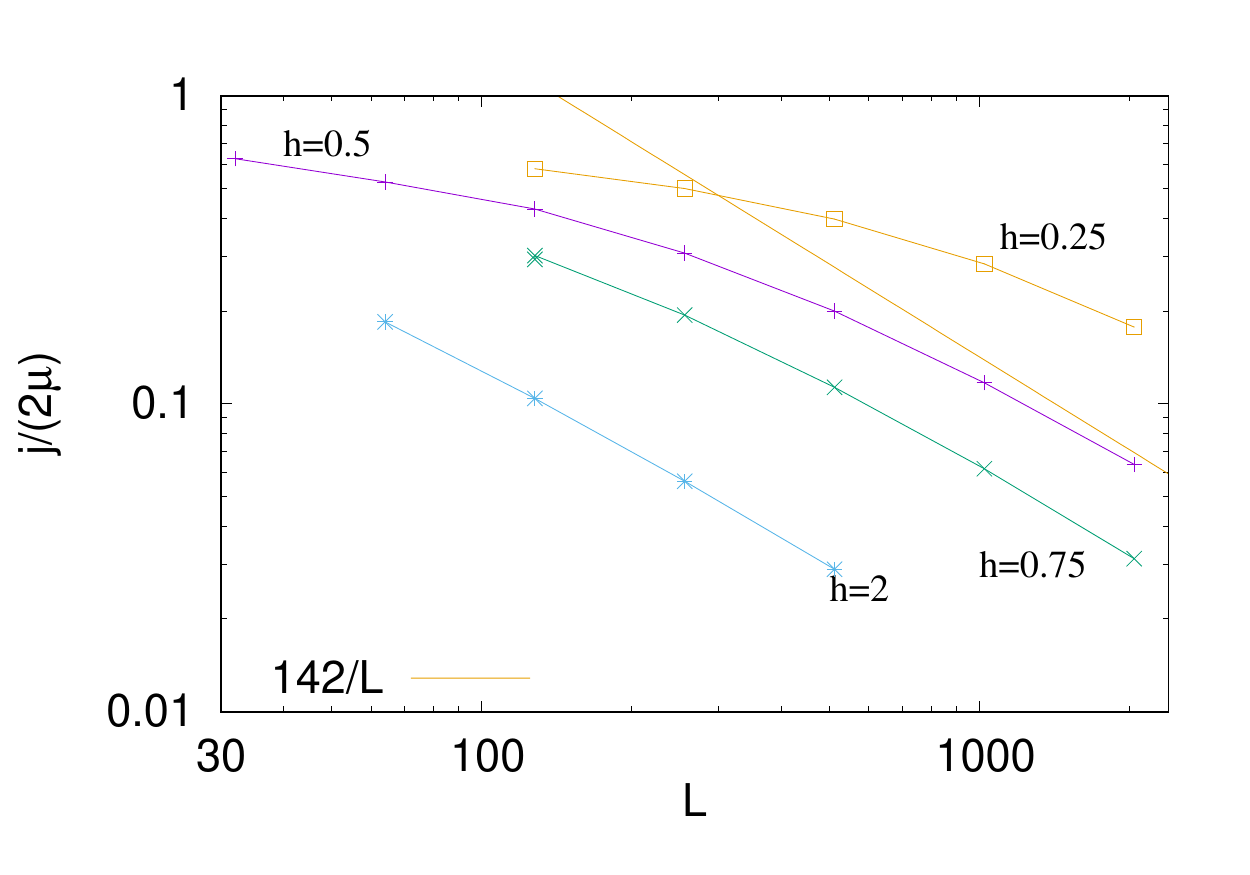}}
\caption{(Color online) Top: Raw data for XXZ with $h=0.5$ from Fig.~\ref{fig:Dodlam}. Bottom: Data used in Fig.~\ref{fig:XXZ}b, where $\lambda=32$, $\Delta=0.6$.}
\label{fig:Dlamall}
\end{figure}

For small $h$ and $\Delta=0.6$, where the unperturbed model is ballistic, the best fitting dependence in Fig.~\ref{fig:Dodh} gives $D \sim 1/h^{1.8}$. The power is not quite $2$, as one would expect from the Fermi's golden rule. We note that something similar has been observed also in the case of disorder with random amplitude at every site in Ref.~\cite{Znidaric16}. In our case we place impurity with the same amplitude $h$ at every $4$ sites. What could play a role is that $\lambda=4$ seems already quite close to the regime of $\lambda \gg 1$, see Fig.~\ref{fig:Dodlam}, where we know from the single-impurity scaling of $R_{\rm single}$ that the power is close to $1.5$.
At $h=0.5$ and $\Delta<1$ we can see in Fig.~\ref{fig:Dlamall} that large $L$ are required in order to reach the asymptotic diffusive spin transport.

\subsection{Single impurity}

We saw in Eq.~(\ref{eq:DR}) that for $\lambda \gg 1$ and $\Delta<1$ we can predict the value of $D$ solely from the single-impurity situation. In the main text we used a fixed $\Delta=0.6$, demonstrating that one has $R_{\rm single}\approx h^{1.5}/1.6$.
\begin{figure}[h!]
\centerline{\includegraphics[width=.75\columnwidth]{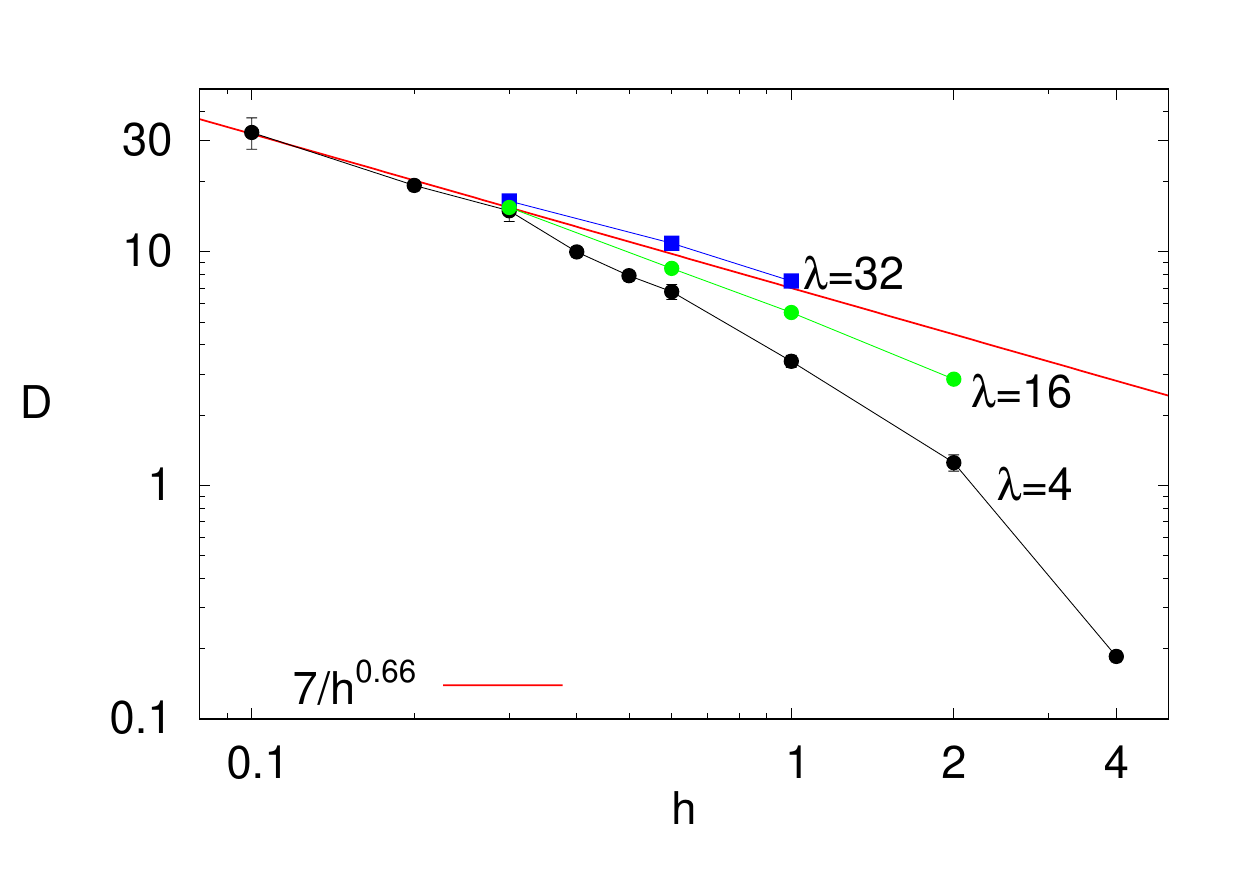}}
\caption{(Color online) Dependence of $D$ on $h$ for the isotropic model, $\Delta=1$, and different values of $\lambda$. Black points for $\lambda=4$ are the same as in Fig.~\ref{fig:XXX}a.}
\label{fig:XXXDodhall}
\end{figure}
Let us remind that at $\Delta=0$ the exact result given in the main text is $R_{\rm single} = h^2/4$ and therefore $D(\Delta=0) \sim \lambda/h^2$. On the other hand, for $\Delta=1$ and intermediate $\lambda=4$ we have seen in Fig.~\ref{fig:XXX}a that for small $h$ the power is around $D \sim 1/h^{0.66}$. In Fig.~\ref{fig:XXXDodhall} we show also data for $\lambda=16$ and $32$. While that data is less precise than the one for $\lambda=4$, and we do not have datapoints at very small $h$, one can nevertheless see that also at large $\lambda$ the behavior is still compatible with $D \sim 1/h^{0.66}$ at small $h$. Summarizing, at large $\lambda$ one has $D \sim 1/h^2$ at $\Delta=0$, and much smaller power at $\Delta=1$, where $D \sim 1/h^{0.66}$. On the other hand at $\Delta=0.6$ we have seen in Fig.~\ref{fig:XXZ}b that $D \sim 1/h^{1.5}$.

\begin{figure}[ht!]
\centerline{\includegraphics[width=.75\columnwidth]{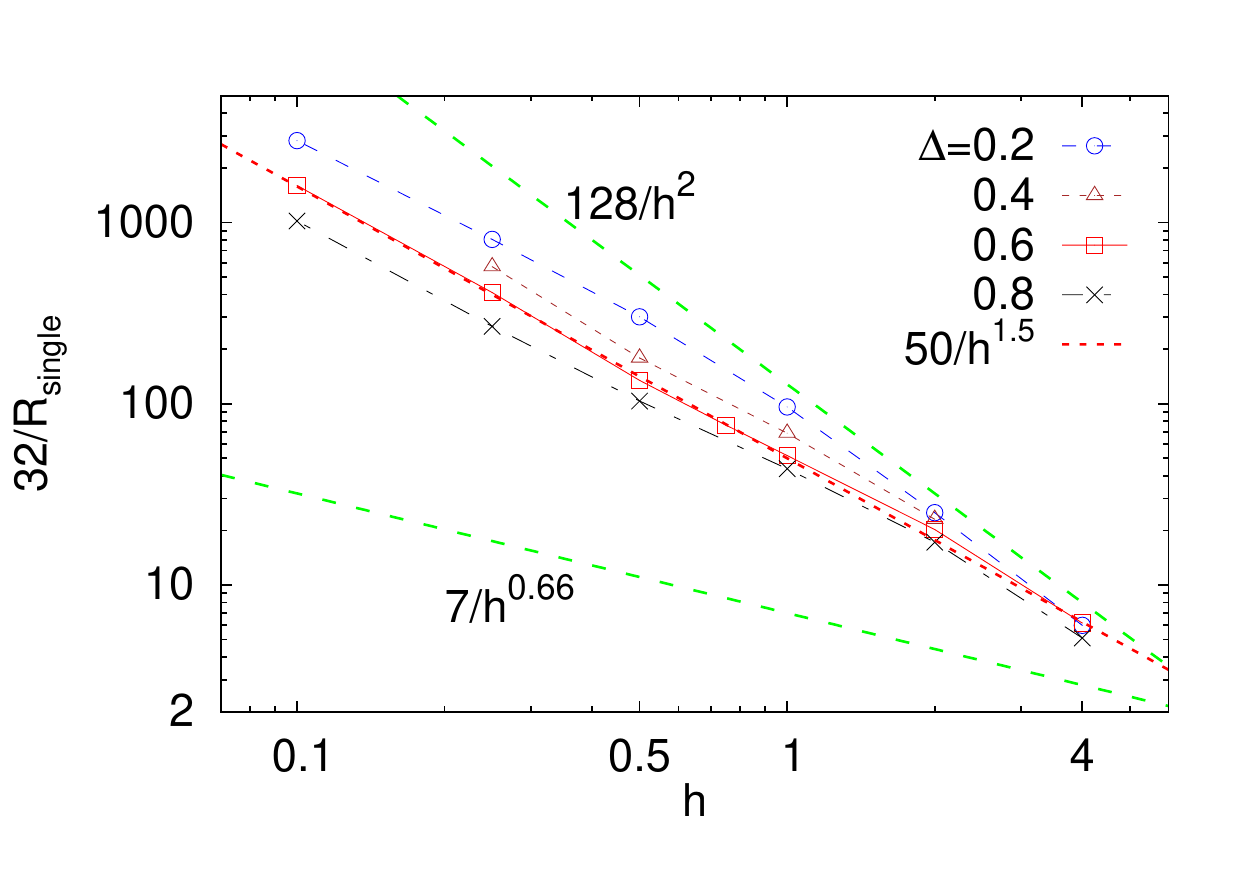}}
\caption{(Color online) Scaling of $R_{\rm single}$ with $h$ for different $\Delta$, all obtained for $L=128$ and using 5 central points to determine $dz$, like in Fig.~\ref{fig:XXZ}c. Red squares is the same data for $\Delta=0.6$ shown in Fig.~\ref{fig:XXZ}b. Two green lines denote dependence $D=4\lambda/h^2$ for $\Delta=0$ (using $\lambda=32$), and $D \approx 7/h^{0.66}$ holding for $\Delta=1$.}
\label{fig:jdmuDall}
\end{figure}

An obvious question is how does the power depend on $\Delta$, is it continuously changing from $2$ to $0.66$ as one changes $\Delta$, or is it discontinuous? The question is not easy to answer as one will have to deal with large finite-size effect at small $\Delta$ (as well as possibly at $\Delta$ close to $1$). A detailed treatement goes beyond the present work, however we nevertheless present some results shedding light on the problem. In Fig.~\ref{fig:jdmuDall} we show single-impurity results for $R_{\rm single}$ and few additional values of $\Delta$. We can see that at all 4 values of $\Delta=0.2,0.4,0.6,0.8$ the power at small $h$ is rather close to $1.5$ (and is clearly distinct from both $2$ at $\Delta=0$ and $\approx 0.66$ at $\Delta=1$). More precisely, the best fitting power at small $h$ is $1.39, 1.55, 1.50, 1.4$ at $\Delta=0.2,0.4,0.6,0.8$, respectively (if 3 central points would be used in determining $ds$ the powers would be $1.7, 1.65, 1.65, 1.5$, respectively). While it is hard to say about the power close to $\Delta=0$ and $\Delta=1$, data is consistent with a discontinuous change in the exponent, i.e., at fixed $\Delta$ and small $h$ the power is different than at $\Delta=0$ or $1$. If this is the case one has an interesting situation where $D(\Delta=1)\sim \sqrt{\lambda}/h^{0.66}$, while $D(\Delta=1-\varepsilon)\sim \lambda/h^a$ with $a \approx 1.5$. This means that for dilute impurities diffusion is very sensitive to whether one is at the isotropic point $\Delta=1$. In the TDL the ratio of $D(\Delta=1^-)/D(\Delta=1)$ will in fact diverge with large $\lambda$ or small $h$. To be concrete, taking $\lambda=32$ and $h=0.5$ the diffusion constant $D$ increases about tenfold as one changes interaction from $\Delta=1$ to $\Delta=0.8$ (Fig.~\ref{fig:jdmuDall}).

\begin{figure}[ht!]
\centerline{\includegraphics[width=.75\columnwidth]{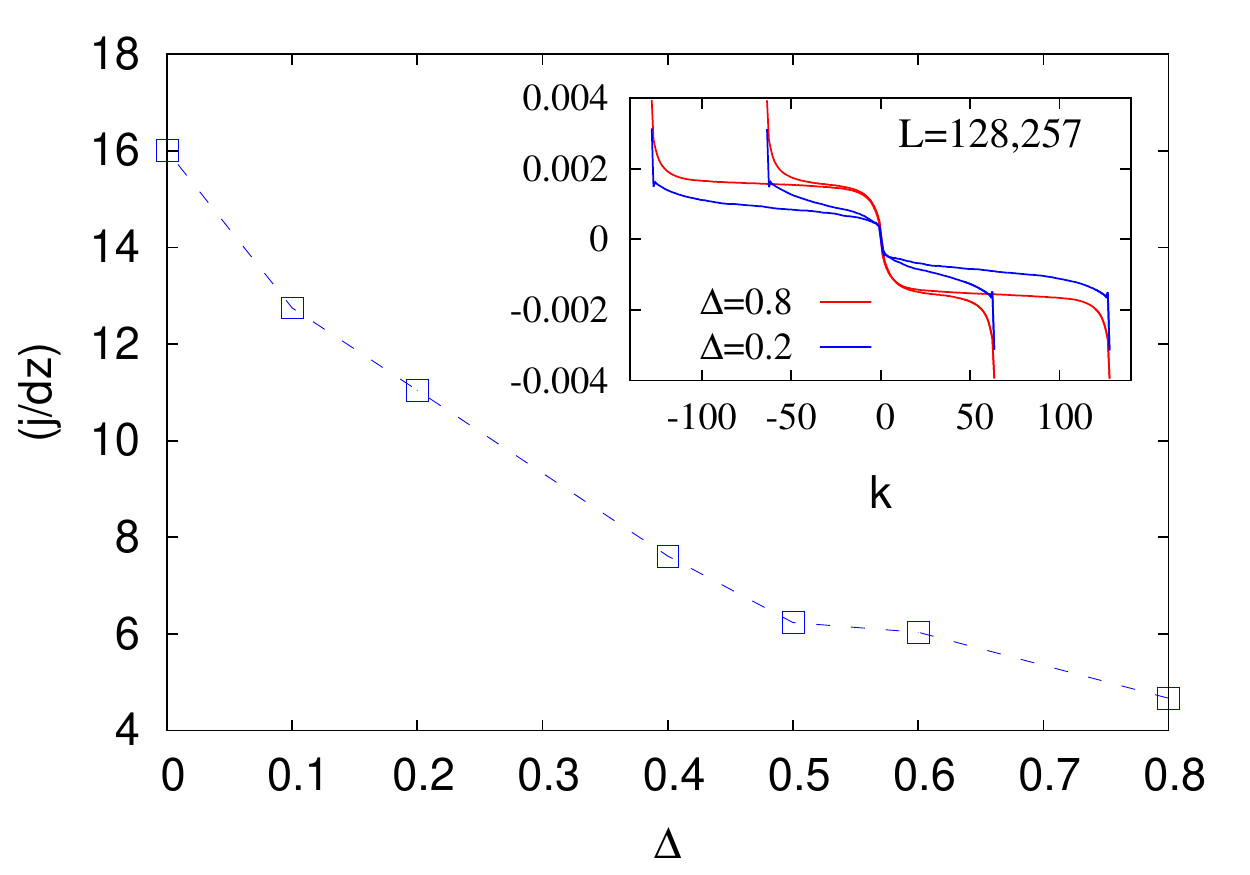}}
\caption{(Color online) Single-impurity resistance. Dependence of $R_{\rm single}=j/dz$ (where $dz$ is magnetization jump at the middle three sites) on $\Delta$ for fixed $h=0.5$. Inset: Profiles for two values of $L$ show that, while at smaller $\Delta$ one indeed has overlap on 3 central sites, at larger $\Delta$ one would have to take $dz$ on more than 3 points (driving strength is $\mu=0.005$).}
\label{fig:jdmuh05}
\end{figure}
The nontrivial power $\approx 1.5$ therefore seems to be changing discontinuously at both $\Delta=0$ and $\Delta=1$ (or at least it changes very rapidly, so that at our smallest $\Delta=0.2$ and largest $\Delta=0.8$ we could not see a smooth dependence). Because we are dealing with large $\lambda$, where there are jumps in the magnetization only at impurities, it must be a consequence of the jump itself scaling as $dz \sim h^{1.5}$ at small $h$ (the current for $h \ll 1$ on the other hand saturates at an $h$ independent value that is the same as for the clean model). Several possible perturbative approaches to obtain the power do not work. One is using a small-$h$ Liouvillian perturbation theory, which would however result in an integer power, like $2$ starting from unperturbed $\Delta=0$. Another way, which also fails, is using the Fermi's golden rule on a noninteracting model with $\Delta=0$, taking periodic impurities as the perturbation potential $V\sim h\sum_r n_{r\lambda}$. The transition rate $1/\tau$ from a single-particle eigenstate $\ket{k}$ to another $\ket{k'}$ is proportional to the matrix element $|\bracket{k'}{V}{k}|^2$. Single-particle eigenstates $\ket{k}$ are plane-waves and we can write their wavenumber as $k=\frac{2\pi}{L}p$ and $k'=\frac{2\pi}{L}p'$, where $p,p'$ are integers. Doing the calculation we get $|\bracket{k'}{V}{k}|^2 \sim \frac{h^2}{\lambda^2} \frac{\sin^2{[\pi(p-p')]}}{[\pi(p-p')]^2}$. We see that (i) if $p-p'$ is really an integer, like for $\Delta=0$, there is no scattering, as it should be. Adding a periodic potential to free particles does not modify the ballistic transport. If one wants to have a scattering that breaks ballistic transport one needs interaction $\Delta \neq 0$, so that momenta are not integers anymore. (ii) Focusing on the prefactor, the scattering rate is $1/\tau \sim h^2$ which is the correct power of the magnetization jump on the single-impurity only if $\Delta=0$ (the case for which we anyway exactly solved the Lindblad equation). (iii) $\tau \sim \lambda^2$ would suggest $D \sim \lambda^2$, which is not correct. The golden-rule factor $1/\lambda^2$ comes about simply due to the norm of $V$; for larger $\lambda$ there are simply less impurities in the system. And because in the Fermi's golden rule we are summing amplitudes of scattering at different sites (different terms in $V$), we get the factor $\lambda^2$. According to Matthiessen's rule one instead has to add rates (probabilities), which then gives the correct scaling $D \sim \lambda$. Note though that neither of the two arguments gives the correct scaling at the isotropic point where $D \sim \sqrt{\lambda}$.

In Fig.~\ref{fig:jdmuh05} we study how $R_{\rm single}$ depends on $\Delta$, this time fixing $h=0.5$. We can see that $R_{\rm single}$ increases by decreasing $\Delta$ (this is also visible in Fig.~\ref{fig:jdmuDall}). Beware that the limit $\Delta=0$ is special in the sense that the model is ballistic even with $h\neq 0$ periodic impurities. In that figure we read $dz$ from the jump at the middle three sites, regardless of $\Delta$. At larger $\Delta$ one should in fact take more than three sites because the width of the magnetization jump at the impurity depends on $\Delta$. For instance, for $\Delta=0.6$ used in the main text we determined that 5 sites is more appropriate. Such an adjustment would slightly lower the curve shown in Fig.~\ref{fig:jdmuh05} at larger $\Delta$.

\subsection{Role of driving and coupling strength}

\begin{figure}[h!]
\centerline{\includegraphics[width=.75\columnwidth]{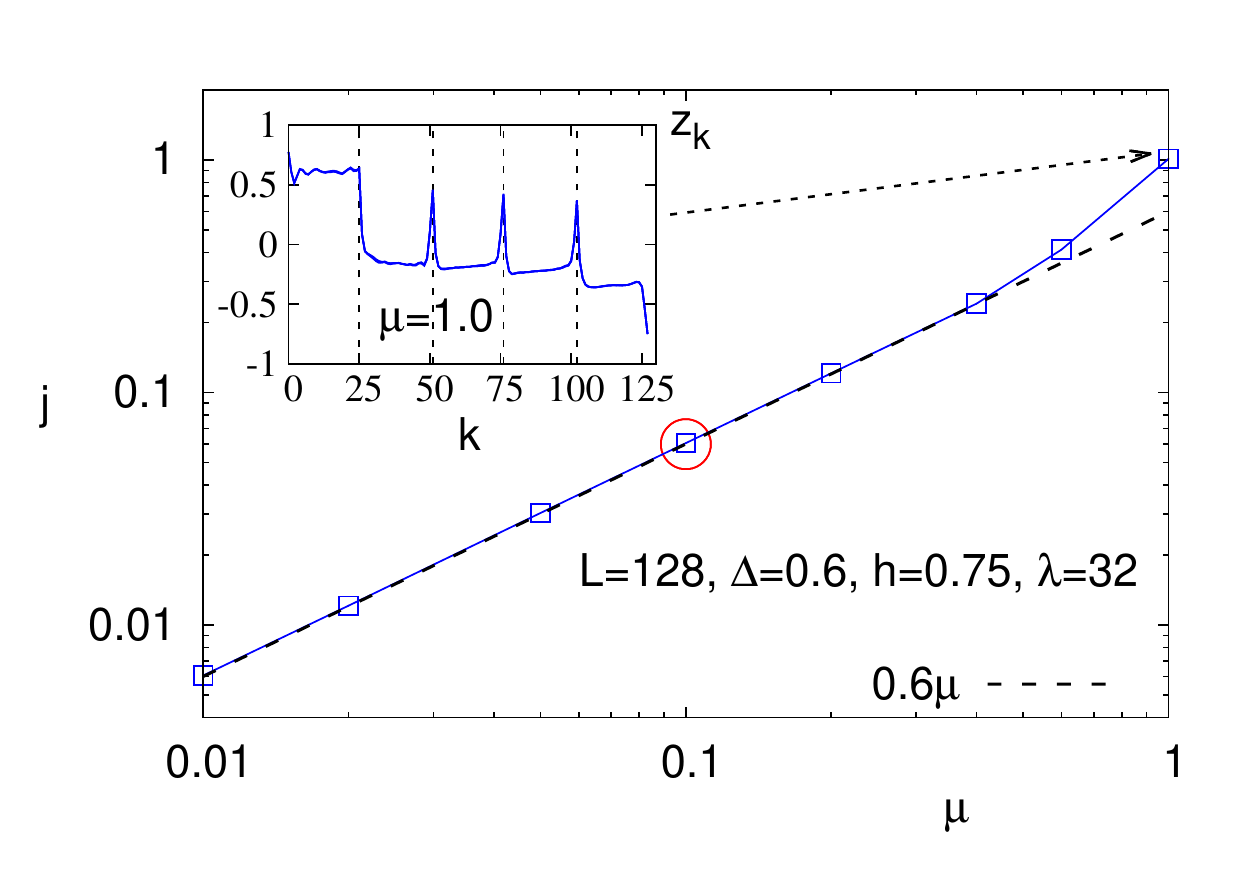}}
\caption{(Color online) Linear dependence of the NESS current $j$ on the driving strength $\mu$, except at very large driving $\mu>0.4$. The circled $\mu=0.1$ is the parameter choice used in the paper (the corresponding profile is shown in Fig.~\ref{fig:profilG3}). Inset: Magnetization profile at the maximal driving $\mu=1$, where one has spikes at impurity positions (4 vertical lines). Parameters are $\lambda=32$, $\Delta=0.6$, $h=0.75$.}
\label{fig:jodmu}
\end{figure}

Lindblad parameter $\mu$ determines how strong the driving is. The linear response regime that we want to probe is defined as a regime where the NESS current is proportional to $\mu$. One therefore has to take a sufficiently small $\mu$ such that this is the case. In Fig.~\ref{fig:jodmu} we show an example of the dependence of current $j$ on $\mu$. We can see that the chosen $\mu=0.1$ used in the rest of the paper is indeed in the linear response regime.

\begin{figure}[ht!]
\centerline{\includegraphics[width=.75\columnwidth]{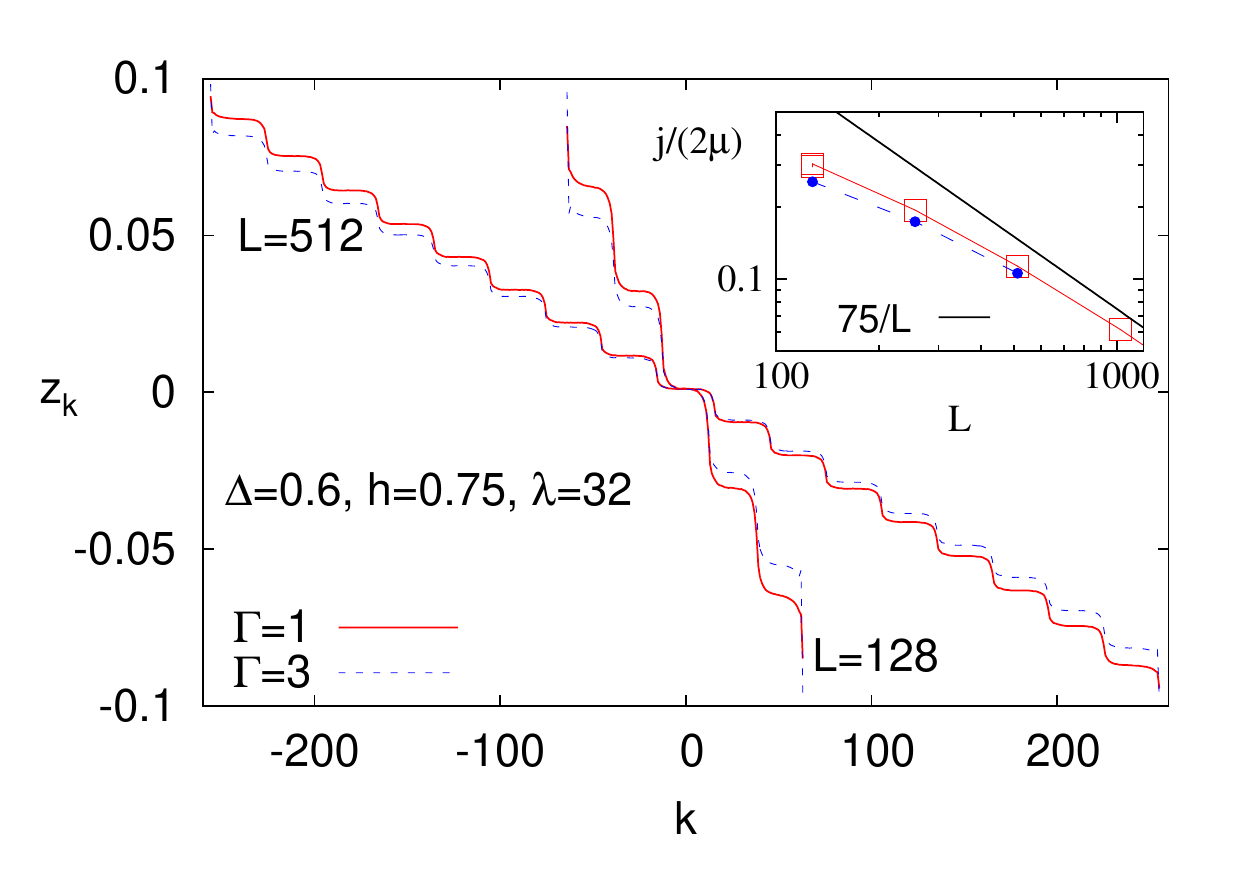}}
\caption{(Color online) Comparing bath coupling strength $\Gamma=1$ used throughout the paper (red) with $\Gamma=3$ (blue). Boundary jumps due to larger boundary resistance are fairly large for $L=128$, however, at larger $L$ the boundary jump gets small. This is in-line with the NESS currents being the same in the TDL (inset). Parameters are $\lambda=32$, $\Delta=0.6$, $h=0.75$, and driving $\mu=0.1$.}
\label{fig:profilG3}
\end{figure}

In all our non-integrable cases we observe diffusion, for which the NESS magnetization profile is (on average) linear. The linear profile follows from $j=-D dz/dx$ as long as magnetization varies little across the chain so that $D$ can be considered a constant, while $j$ is constant in the NESS due to the continuity equation. If the model is ballistic, as e.g. it is without impurities for $\Delta<1$, the profile is flat in the zero-resistance bulk. In short, the local magnetization gradient is an indicator of the local resistance -- in places of high resistance the gradient is large, in places of low resistance it is small. This also explains a step-like profile (which is still linear on average) at large $\lambda$ (e.g. Fig.~\ref{fig:XXZ}a or Fig.~\ref{fig:profilG3}). 

In Fig.~\ref{fig:jodmu} we see interesting strong spikes in the profile at very large $\mu$ outside of the linear response regime. At the maximal driving $\mu=1$ one allows transport of spin only in one direction (at the left edge only $L_1 \propto \sigma_0^+$ acts, while at the right edge only $L_4 \propto \sigma_{L-1}^-$). This apparently increases the average current $j$ compared to its linear response value, making the profile, apart from spikes, also rather flat (ballistic) in the bulk. This interesting detail needs further attention to see whether it persists in the TDL.

Finally, in Fig.~\ref{fig:profilG3} we check that asymptotically at large $L$ the value of $D$ does not depend in the Lindblad coupling strength $\Gamma$. In all other simulation we used $\Gamma=1$, while here we also show $\Gamma=3$. The value of $\Gamma$ essentially determines the boundary resistance, and therefore only influences the magnetization jump at the boundary $dz_{\rm boundary} \sim R_{\rm boundary}j$. If one has diffusion where the current scales as $j \sim 1/L$ in the TDL, the size of this boundary jump will go to zero and will not affect $D$. This is in line with analytical arguments in Ref.~\cite{nessKubo}. In the thermodynamic limit the coupling strength $\Gamma$ can play a role only if the transport is ballistic, that is if $z=1$. As soon as one is sub-ballistic ($z>1$) $\Gamma$ does not matter in the TDL (in practice, a large or a small $\Gamma$ could be a numerical nuisance due to large $R_{\rm boundary}$ and slower convergence with $L$).


\begin{thebibliography}{99}
\bibitem{foot1} We try to place $M$ equidistant impurities; for finite $L$ the distance between few of them can differ from $\lambda$ by $\pm 1$, which is, however, irrelevant in the TDL.
\end{thebibliography}
\end{document}